\documentclass[preprint,authoryear]{elsarticle}
\usepackage{amssymb,amsmath,epsfig}
\usepackage{paralist}
\usepackage{algpseudocode}
\usepackage{algorithm}
\algtext*{EndWhile}
\algtext*{EndIf}
\algtext*{EndFor}
\makeatletter  
\renewcommand{\ALG@name}{Table}
\makeatother   
\algnewcommand{\Initialize}[1]{%
  \State \textbf{Initialize}
  \Statex \hspace*{\algorithmicindent}\parbox[t]{.8\linewidth}{\raggedright #1}
}

\def\av#1{\left\langle#1\right\rangle}
\def\to{\rightarrow}
\def\Var{{\rm Var}\,}

\def\di{\partial}

\def\tpsi{\tilde{\psi}}
\def\tP{\tilde{P}}
\def\teta{\tilde{\eta}}
\def\tphi{\tilde{\phi}}
\def\tQ{\tilde{Q}}
\def\smc{{\scriptscriptstyle \textrm{SMC}}}
\def\smcp{{\scriptscriptstyle \textrm{SMC'}}}
\def\smcpseg{{\scriptscriptstyle \textrm{SMC'}},\textrm{seg}}

\begin{document}


\title{A Renewal Theory Approach to IBD Sharing}

\author[cu]{Shai Carmi\corref{cor}}
\ead{scarmi@cs.columbia.edu}
\author[har]{Peter R. Wilton}
\author[har]{John Wakeley}
\author[cu]{Itsik Pe'er}

\address[cu]{Department of Computer Science, Columbia University, New York, NY, 10027, USA}
\address[har]{Department of Organismic and Evolutionary Biology, Harvard University, Cambridge, MA, 02138, USA}
\cortext[cor]{Corresponding author}

\begin{abstract}
A long genomic segment inherited by a pair of individuals from a single, recent common ancestor is said to be \emph{identical-by-descent} (IBD). Shared IBD segments have numerous applications in genetics, from demographic inference to phasing, imputation, pedigree reconstruction, and disease mapping. Here, we provide a theoretical analysis of IBD sharing under Markovian approximations of the coalescent with recombination. We describe a general framework for the IBD process along the chromosome under the Markovian models (SMC/SMC'), as well as introduce and justify a new model, which we term the \emph{renewal approximation}, under which lengths of successive segments are independent. Then, considering the infinite-chromosome limit of the IBD process, we recover previous results (for SMC) and derive new results (for SMC') for the mean number of shared segments longer than a cutoff and the fraction of the chromosome found in such segments. We then use renewal theory to derive an expression (in Laplace space) for the distribution of the number of shared segments and demonstrate implications for demographic inference. We also compute (again, in Laplace space) the distribution of the fraction of the chromosome in shared segments, from which we obtain explicit expressions for the first two moments. Finally, we generalize all results to populations with a variable effective size.
\end{abstract}


\maketitle

\section{Introduction}

IBD sharing of a genomic segment between a pair of individuals is traditionally defined in terms of recent co-ancestry, no more remote than some time depth $t$ \citep{ThompsonReview}. In population samples, the time of the common ancestor is unknown, and in practice, IBD segments are often identified as long stretches that are nearly or fully identical-by-state (IBS), to an extent distinguishable from population-level LD. The decision whether a segment is called IBD is either rule-based (e.g., using a certain length cutoff) or model-based, using an underlying hidden Markov model for the IBD state \citep{ThompsonReview}. In this paper, we define an IBD segment shared between two chromosomes as the maximal sequence over which the chromosomes have the same most recent common ancestor (MRCA). Recent mutations (or genotyping errors) separating the two sequences do not disqualify the segment from being IBD. On the other hand, we require the segment to be longer than an (arbitrary) cutoff $m$. This definition enables a theoretical treatment, while largely capturing the way in which some methods (and, for sufficiently large $m$, virtually all methods) discover IBD segments in real data.

Much attention has recently been devoted to efficient algorithms for IBD detection in large samples (e.g., \cite{Plink,Gusev2009,Browning_FastIBD,Brown_HMMIBD,Browning_RefinedIBD}, to give a few examples). Detected segments have found numerous applications, for example, characterization of relationships between populations \citep{Atzmon2010,BrayPNAS2010,RomaHistory,FrenchCanada_IBD,HennNorthAfrica2,RalphCoop}, detection of positive selection \citep{HanAbney_selection}, estimation of heritability \citep{Browning_heritability}, mapping haplotypes associated with a trait \citep{DASH,Browning_IBDMapping2012,IBDMapping_MS}, phasing and imputation \citep{Kong_IBD,PhasingIBD_Durbin}, and pedigree reconstruction \citep{Huff_ERSA,Henn2012}. See \cite{IBD_Review_Browning} and \cite{ThompsonReview} for up-to-date reviews. 

In parallel, theory has been developed for the expected amount of IBD sharing in model populations, with implications for demographic inference. \cite{Palamara2012} and \cite{Palamara_migrations} computed, under the coalescent and for complex demographies, the moments of the fraction of the chromosome found in shared segments of a given length. \cite{Palamara2012} and \cite{HyperSharingGenetics} then approximated the distribution of this quantity, assuming a Poisson distribution for the number of segments (see also \cite{Huff_ERSA}). \cite{RalphCoop} computed the expected number of shared segments of a certain length given an arbitrary demographic history. However, certain theoretical problems of interest have remained open.

Here, we introduce a general framework for the analysis of the IBD process along the chromosome, based on a renewal approximation. Renewal theory is the study of processes in which events are separated by independent waiting times, and where each waiting period or event may be associated with a value \citep{KarlinTaylor}. Under certain conditions, consecutive shared segments along the chromosome can be approximated as independent. Then, interpreting segments with shared ancestry as waiting times, renewal theory can be applied to compute the distribution of the number of and the total amount of genetic material covered by segments of a certain length. 

A renewal approach to the IBD process has been considered in the past (e.g., \cite{Stam1980,Chapman2003}, with initial contributions already by \cite{Fisher1954}), in a model where the population has been recently founded by individuals of heterogeneous genetic types. Alternatively, in those works, IBD is defined with respect to a given time depth \citep{ThompsonReview}. The IBD segment lengths were either assumed exponential or fitted. In contrast, we consider a model that can be applied without reference to a particular time point. In our model, two chromosomes can trace their common ancestor, at each locus, to any time in the past, and IBD segments are defined with respect to a length cutoff.

According to our renewal approximation for a pair of chromosomes, the time to the common ancestor is drawn, at a recombination event, independently of the previous time and from a position-independent stationary distribution. The distribution has been derived for the pairwise Sequentially Markov Coalescent (SMC) by \cite{LiDurbin2011}, and we derive it here for the more accurate, yet tractable SMC' model \citep{SMCprime}. Under this approximation, the distribution of segment lengths emerges naturally. Using renewal theory, we are then able to derive new results, such as the distribution of the number of shared segments, as well as recover previous results as special cases.  

Our results are organized as follows. In section \ref{sect_ibd_process}, we introduce the renewal approximation in the context of successively simplified approximations of the coalescent with recombination. We then describe the IBD process under the different models and present numerical evidence to justify the renewal approach. In section \ref{sect_ibd_infinite_chr}, we show how simple quantities, such as the mean number of shared segments and the mean fraction of the chromosome in shared segments, emerge naturally from our definition of the IBD process by taking the infinite-chromosome limit. Specifically, we recover previous results for SMC and obtain new results for SMC'. In section \ref{sect_finite_chr}, we derive results for finite chromosomes. Specifically, we derive an expression, in Laplace space, for the distribution of the number of shared segments and consider implications for demographic inference. Additionally, we derive, again in Laplace space, the distribution of the fraction of the chromosome found in shared segments, from which we obtain explicit expressions for the first two moments, recovering and extending previous results. Finally, in section \ref{sect_ibd_varsize}, we generalize our results to populations with variable size. We summarize and discuss the results in section \ref{sect_discussion}.

\section{The IBD process}

\label{sect_ibd_process}

\subsection{Overview of the coalescent with recombination and its Markovian approximations}
\label{sect_ibd_models}

\begin{figure}[!t]
\includegraphics[width=8cm,height=5.5cm]{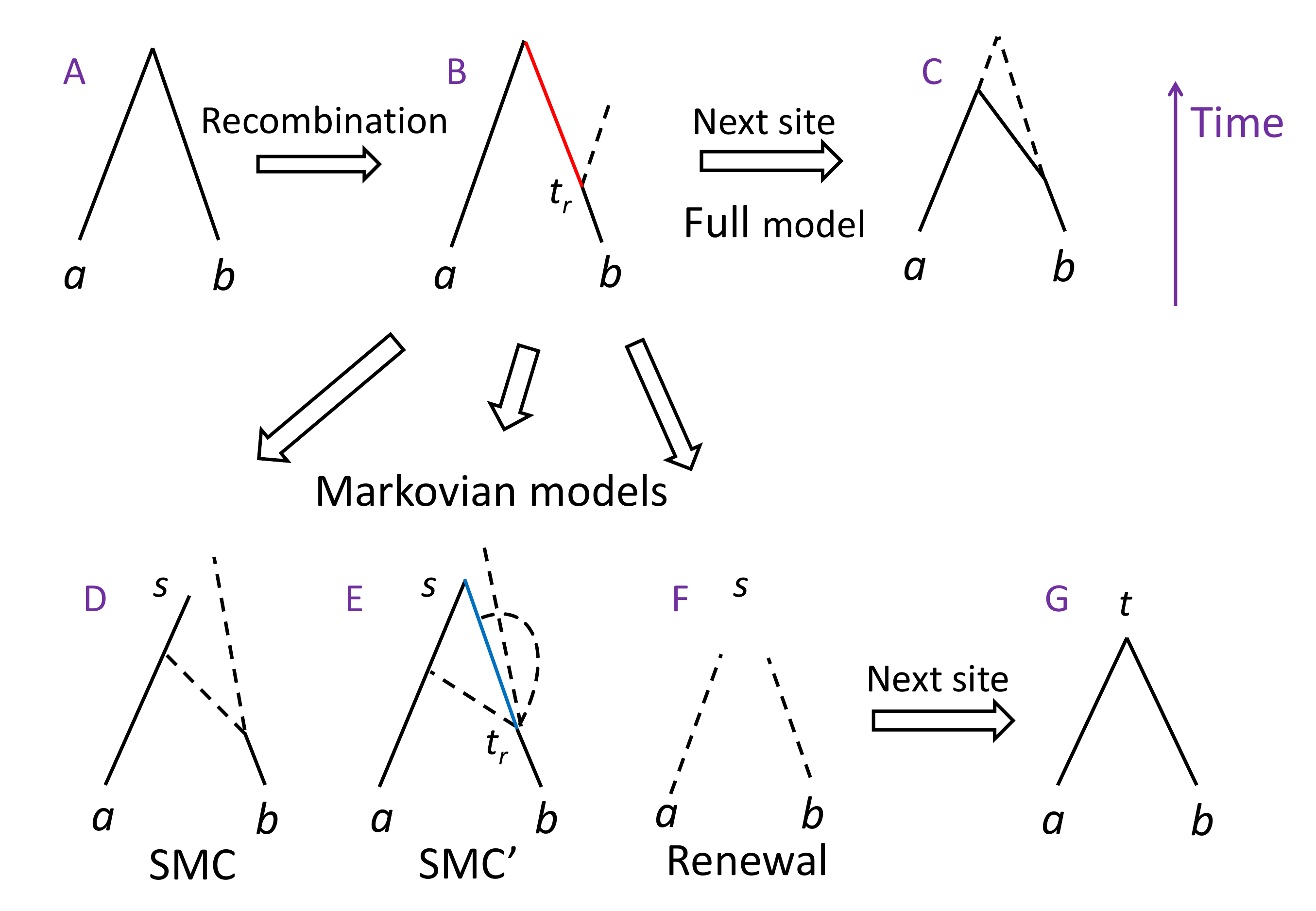}
\caption{An illustration of the coalescent with recombination for two chromosomes, and the associated Markovian approximations. Part \textbf{A} shows the coalescent tree at a random site. The two extant chromosomes are denoted $a$ and $b$. Part \textbf{B} is indicating a recombination event occurring at time $t_r$. The old branch connecting the breakpoint and the MRCA is colored red, and the branching lineage is shown as a dashed line. Under the full model of the coalescent with recombination (the ARG; \cite{WiufHein1999}; \textbf{C}), the new lineage can coalesce with any branch in the existing tree (in this example, earlier than the previous TMRCA), and both the old lineage (which is not ancestral to the sample anymore) and the new lineage are carried over to the next site. The `marginal' tree at the new site is shown in solid lines; the remainder of the ARG is in dashed lines. The Markovian approximations are presented in parts \textbf{D}-\textbf{G}, where the current TMRCA is denoted as $s$ and the new as $t$. In SMC (\cite{McVean2005}; \textbf{D}), the old branch (red in \textbf{B}) is deleted, and the branching lineage can coalesce only with the lineage corresponding to the other chromosome (either earlier or later than the previous TMRCA; corresponding to the two dashed lines). In SMC' (\cite{SMCprime}; \textbf{E}), the branching lineage can coalesce with the old branch (blue), but that branch is deleted once the new tree is formed. Under the renewal approximation (\textbf{F}), the new tree height is drawn independently of the previous tree height. In all Markovian approximations, the new tree (\textbf{G}) contains only the lineages ancestral to the sample at that position.} \label{fig_arg}
\end{figure}

We consider a sample of two chromosomes of length $L$ (Morgans) in a population of a constant effective size $N$ (haploid chromosomes) and with recombination modeled as a Poisson process along the chromosome. The ancestral process can be described by the coalescent with recombination \citep{Hudson1983,GriffithsMarjoram1997}. In that model, looking backwards in time, lineages can either coalesce (at rate 1 per pair of lineages, when the time is scaled by $N$) or recombine at a random position along the chromosome (split into two, at rate $\rho=2Nr$, where $r$ is the recombination probability per generation). The resulting structure is called the \emph{ancestral recombination graph} (ARG). \cite{WiufHein1999} described an alternative but equivalent formulation, where the ARG is obtained by walking along the chromosome. In that model, a coalescent tree is first formed at the leftmost end of the chromosome (Figure \ref{fig_arg}A). Recombination then occurs at a genetic distance distributed exponentially with rate equal to the total branch length of the tree; the position of the breakpoint ($t_r$) is randomly and uniformly distributed along the tree (Figure \ref{fig_arg}B). The branching lineage then coalesces with any of the existing branches of the ARG, and the process is repeated until reaching the end of the chromosome (Figure \ref{fig_arg}C). The model is non-Markovian, in the sense that the tree formed at a given position depends on all preceding trees.

\cite{McVean2005} proposed a Markovian approximation to the coalescent with recombination (the \emph{Sequentially Markov Coalescent}, or SMC). At each recombination event in SMC, the branch leading from the breakpoint to the most recent common ancestor (MRCA) is deleted, and the branching lineage is allowed to coalesce only with the lineage ancestral to the other individual (Figure \ref{fig_arg}D,G). Once the MRCA is reached, the process is continued with the newly formed tree. \cite{SMCprime} suggested a more accurate approximation, called SMC', in which the branching lineage is allowed to coalesce with the branch it had split from, but once the tree has formed, any branch not ancestral to the sample is again deleted (Figure \ref{fig_arg}E,G). See \cite{Hobolth2014} for the joint distribution of tree heights for two sequences at two loci under the ARG and the Markovian approximations.


We propose the \emph{renewal approximation}, which is a further simplification of SMC. According to our approximation, at a recombination event, the new tree height is drawn, \emph{independently} of the previous tree height, from the stationary distribution of tree heights under SMC (Figure \ref{fig_arg}F,G). The stationary distribution was derived by \cite{LiDurbin2011} (see the next section). While the independence assumption is strong, the fact that we use the SMC stationary distribution guarantees that for sufficiently long sequences (see simulations in section \ref{sect_intro_simulations}), the statistical properties of SMC and the renewal process are similar.

In the following subsections, we define the IBD process under the three models: SMC, renewal, and SMC' (Tables \ref{process_smc}-\ref{process_smcprime}, respectively).

\subsection{The IBD process under SMC}

\label{sect_ibd_process_intro}

\begin{figure}
\includegraphics[width=8cm,height=6cm]{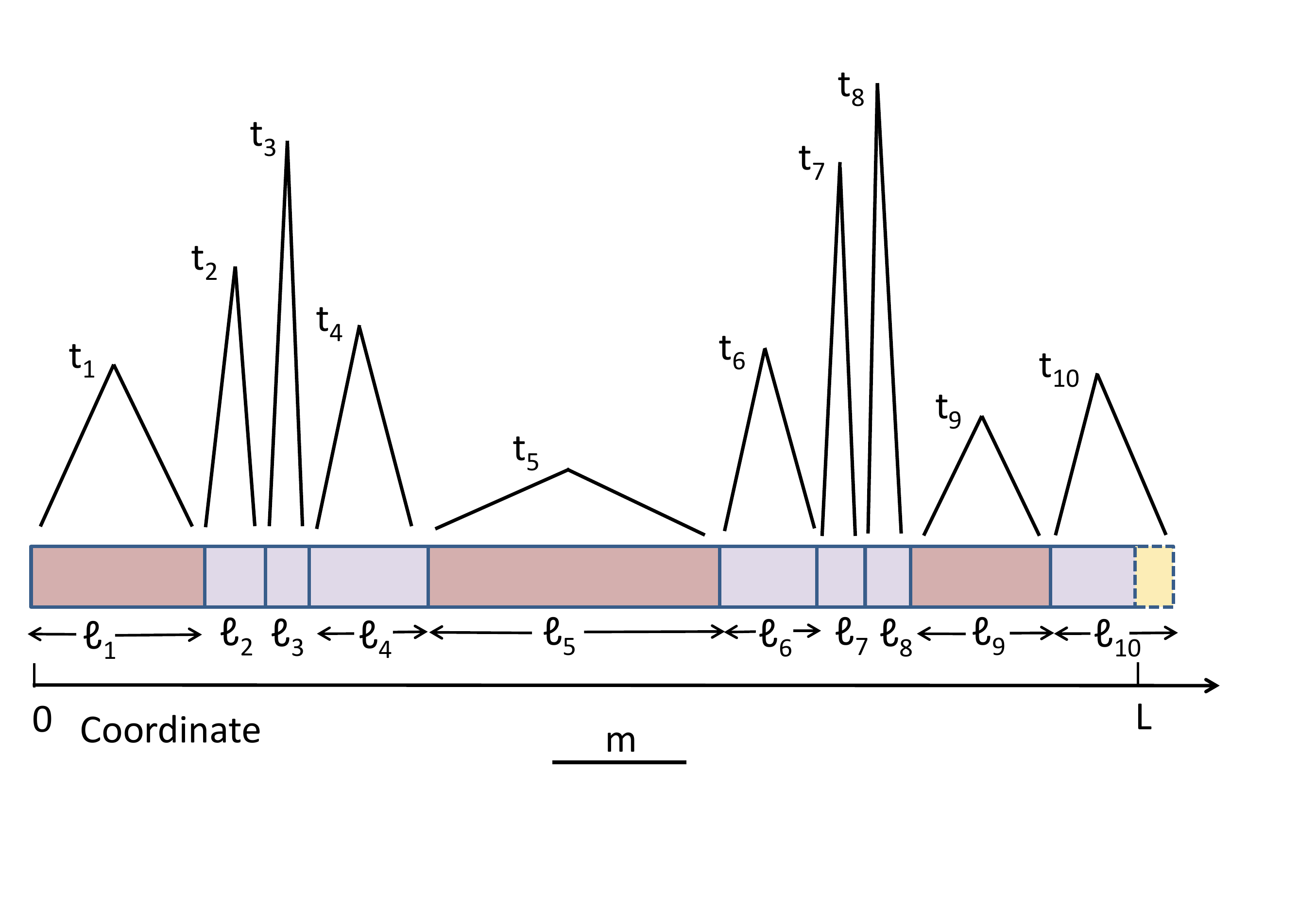}
\caption{An illustration of the IBD process along the chromosome under SMC. Segments are broken by recombination events (vertical bars). The TMRCA is shown on top of each segment. Given a TMRCA $t_i$ at segment $i$, the segment length, $\ell_i$, is distributed exponentially with rate $2Nt_i$, and the TMRCA at the next segment, $t_{i+1}$, is distributed according to Eq. \eqref{eq_smc_transitions}. The minimal segment length, $m$, is shown as a horizontal bar under the chromosome. Segments longer than $m$ are shown in dark pink. In this example, there are three such segments; hence $n_m=3$ and the fraction of the chromosome in shared segments is $f_m=(\ell_1+\ell_5+\ell_9)/L$. Segments shorter than $m$ are in light pink. The last segment exceeds the chromosome length; the excess length (yellow) is ignored.} \label{fig_ibd_process_renewal}
\end{figure}

Recently, \cite{LiDurbin2011} derived the probability density function (PDF) of the tree height for a pair of chromosomes (equivalently, time to MRCA or TMRCA; and scaled by $N$) at a recombination site, given the TMRCA of the preceding tree. The result is given in their supplementary Eq. (6),
\begin{equation}
\label{eq_smc_transitions}
q_{\smc}(t|s)=\begin{cases}
\frac{1}{s}\left(1-e^{-t}\right) & t<s, \\
\frac{1}{s}e^{-(t-s)}\left(1-e^{-s}\right) & t>s,
\end{cases}
\end{equation}
where $s$ and $t$ are the previous and new TMRCA, respectively. Note that $t\ne s$ by definition and that $q_{\smc}(t|s)$ is normalized. At a recombination site, and for a given new tree height $t$, the sequence length to the next recombination event is distributed exponentially with rate $2Nt$, the total branch length of the tree (in generations; \cite{WiufHein1999}). The sequence between recombination sites is a \emph{shared segment}, because the common ancestor of the two chromosomes is fixed throughout the segment. In SMC, the MRCA necessarily changes at recombination sites; therefore, segments are terminated by recombination events. With these preliminaries, and imposing a minimal segment length cutoff, $m$, we define in Table \ref{process_smc} the IBD process along the chromosome (see also Figure \ref{fig_ibd_process_renewal}).

\begin{algorithm}[H]
\caption{The IBD process under SMC}
\label{process_smc}
\begin{algorithmic}[1]
\State \textbf{Initialize}
\State \hspace{1em} $x\gets 0$ \Comment \footnotesize The position along the chromosome \normalsize
\State \hspace{1em} $n_m\gets 0$ \Comment \footnotesize The number of shared segments longer than $m$ \normalsize
\State \hspace{1em} $f_m\gets 0$ \Comment \footnotesize The fraction of the chromosome in shared segments longer than $m$ \normalsize
\State \hspace{1em} Draw TMRCA: $t\sim\textrm{Exp}(1)$
\While {$x<L$}
\State \label{item_smc_start_loop} Draw segment length: $\ell \sim \textrm{Exp}(2Nt)$
\If {$(x+\ell)>L$} \Comment \footnotesize If the new position exceeds the chromosome length \normalsize \label{item_renewal_check_boundary1} \State \label{item_renewal_check_boundary2} $\ell \gets (L-x)$ \EndIf
\If {$\ell>m$} \Comment \footnotesize The segment is longer than the cutoff \normalsize
\State $n_m\gets n_m+1$ 
\State $f_m\gets f_m+\ell/L$ 
\EndIf
\State $s\leftarrow t$ 
\State \label{item_smc_draw_t} Draw new TMRCA $t$ with PDF $q_{\smc}(t|s)$ (Eq. \eqref{eq_smc_transitions})
\State $x\gets x+\ell$
\EndWhile
\end{algorithmic}
\end{algorithm}


Steps \ref{item_renewal_check_boundary1} and \ref{item_renewal_check_boundary2} are needed in case the new position exceeds the chromosome length. In simulations, step \ref{item_smc_draw_t} is implemented by drawing a random recombination time, $t_r$, uniform in $[0,s]$, and then a random coalescence time $t_c$, exponential with rate 1. The new TMRCA is then set to $t\leftarrow t_r+t_c$ (Figure \ref{fig_arg}D).

\subsection{The IBD process under the renewal approximation to SMC}

\label{sect_ibd_process_renewal}

Eq. \eqref{eq_smc_transitions} for $q_{\smc}(t|s)$ can be interpreted as the transition probability for a Markov chain whose states are the tree heights at successive recombination sites. \cite{LiDurbin2011}, who derived Eq. \eqref{eq_smc_transitions}, further computed the stationary distribution of the chain,
\begin{equation}
\label{eq_stat_pdf}
\pi_{\infty}^{\smc}(t)=te^{-t}.
\end{equation}
Note that this stationary distribution is not the same as the `marginal' coalescence distribution, $P_c(t) = e^{-t}$, which would apply to the tree height at a pre-specified site, such as the end of a chromosome \citep{WiufHein1999}, or to a randomly chosen site. In fact, $\pi_{\infty}^{\smc}(t)$ is identical to the distribution at a site conditional on a recombination event having occurred at that site when the recombination rate per site is very small. It thus has mean equal to 2 (\cite{GriffithsMarjoram1996}, Eq. (9)), as for example is the case for tree heights around rare insertions in the human genome \citep{HuffEtAl2010}. In other words, $\pi_{\infty}^{\smc}(t)$, may be interpreted as the PDF of the TMRCA of a randomly chosen segment (rather than site).

\begin{figure}
\includegraphics[width=7.5cm,height=5.5cm]{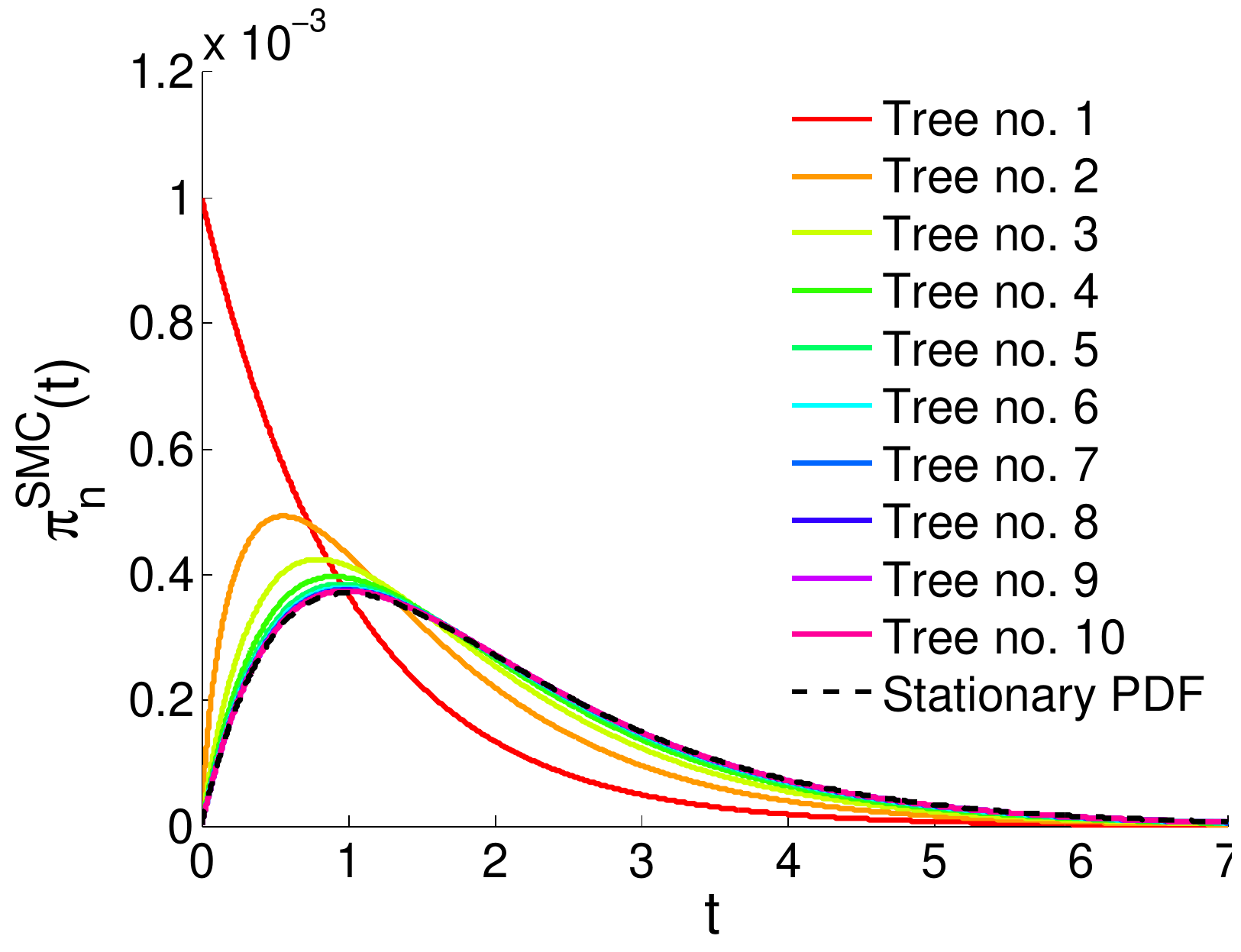}
\caption{Convergence of the distribution of tree heights in the SMC model. The first tree is distributed as $e^{-t}$, according to the standard coalescent. Subsequent trees are distributed according to Eqs. \eqref{eq_smc_transitions} and \eqref{eq_successive_trees}. The integrals were solved numerically. The stationary PDF (dashed line; Eq. \eqref{eq_stat_pdf}) is reached quickly.} \label{fig_stat_pdf}
\end{figure}

To test the convergence to the stationary distribution, we numerically computed the PDFs of successive tree heights, as follows,
\begin{align}
\label{eq_successive_trees}
& \pi_{1}^{\smc}(t)=e^{-t}, \nonumber \\
& \pi_{n+1}^{\smc}(t)=\int_0^{\infty}q_{\smc}(t|s)\pi_{n}^{\smc}(s)ds\;\;;\; n\geq 1.
\end{align}
The resulting PDFs for the first 10 trees are shown in Figure \ref{fig_stat_pdf}, demonstrating fast convergence to the stationary PDF (Eq. \eqref{eq_stat_pdf}). For typical (human) parameters ($N\approx 10^4$, $L\approx 1$ Morgan), the average number of recombination events along the chromosome is $2NL\sim 10^4 \gg 1$ \citep{GriffithsMarjoram1997}. Therefore, the vast majority of trees are expected to have the stationary PDF. 

Using the stationary PDF, segment lengths are therefore distributed as (see also \cite{LiDurbin2011} and \cite{Palamara2012})
\begin{equation}
\label{eq_seg_len_smc}
\psi_{\smc}(\ell)=\int_0^{\infty}\pi_{\infty}^{\smc}(t)\cdot 2Nte^{-2Nt\ell}dt
=\frac{4N}{(1+2N\ell)^3}.
\end{equation}
The mean segment length is $\av{\ell}_{\smc}=1/2N$, but no higher moments exist. The distribution of $\rho=2N\ell$, the scaled recombination rate, is $\psi_{\smc}(\rho)=2/(1+\rho)^3$, which is, as expected, independent of $N$ (a property that holds generally; see section \ref{sect_ibd_varsize}). 

Having the distribution of segment lengths, we can now invoke the assumption of independence between successive segments and define the IBD process in the renewal approximation (Table \ref{process_renewal}). To generate numbers from $\psi_{\smc}(\ell)$, we used the transformation method: let $u$ be uniform in $[0,1]$; we set $\ell=\left(1-\sqrt{u}\right)/\left(2N\sqrt{u}\right)$. 

\begin{algorithm}[H]
\caption{The IBD process under the renewal approximation to SMC}
\label{process_renewal}
\begin{algorithmic}[1]
\State \textbf{Initialize}
\State \hspace{1em} As in Table \ref{process_smc}
\While {$x<L$}
\State Draw segment length $\ell$ with PDF $\psi_{\smc}(\ell)$ (Eq. \eqref{eq_seg_len_smc}) \If {$(x+\ell)>L$} \State $\ell \gets (L-x)$ \EndIf
\If {$\ell>m$} 
\State $n_m\gets n_m+1$ 
\State $f_m\gets f_m+\ell/L$ 
\EndIf
\State $x\gets x+\ell$
\EndWhile
\end{algorithmic}
\end{algorithm}



\subsection{The IBD process under SMC'}

\label{sect_ibd_process_smcprime}

In SMC', the PDF of the new TMRCA, $t$, given the previous TMRCA, $s$, is given by (see also \cite{Zheng2014})
\begin{equation}
\label{eq_transition_pdf_smcprime_raw}
q_{\smcp}(t|s) = \begin{cases}
\int_0^s\frac{1}{s}\left[\int_{t_r}^{s}e^{-2(t_c-t_r)}dt_c\right]dt_r & t=s, \\
\int_0^t\frac{1}{s}e^{-2(t-t_r)}dt_r & t<s, \\
\left[\int_0^s\frac{1}{s}e^{-2(s-t_r)}dt_r\right] e^{-(t-s)} & t>s.
\end{cases}
\end{equation}
To understand Eq. \eqref{eq_transition_pdf_smcprime_raw}, consider how the new TMRCA, $t$, is drawn in simulations. First, a random recombination time, $t_r$, is drawn uniformly in $[0,s]$, as in SMC. But then, the random coalescence time, $t_c$, is drawn from an exponential distribution with rate 2, since the branching lineage can coalesce with either the other chromosome or the lineage it had branched from (Figure \ref{fig_arg}E). If $t_r+t_c<s$, the new TMRCA is set to either $t\leftarrow s$ (coalescence with the lineage it had branched from) or $t\leftarrow t_r+t_c$ (coalescence with the other chromosome) with probability 1/2 each. If $t_r+t_c>s$, a new coalescence time, $\tau_c$, is drawn from an exponential distribution with rate 1 (since after time $s$, there is only one other lineage), and the new TMRCA is set to $t\leftarrow s+\tau_c$. The upper limit of the integral for $t<s$ is $t$, not $s$, since the recombination time, $t_r$, cannot be greater than the new coalescence time, $t$. For the case $t=s$, the density is implicitly assumed to be multiplied by Dirac's delta function ($\delta(t-s)$), omitted for notational simplicity. The integrals in Eq. \eqref{eq_transition_pdf_smcprime_raw} can be solved, yielding
\begin{equation}
\label{eq_transition_pdf_smcprime_final}
q_{\smcp}(t|s) = \begin{cases}
\frac{2t+e^{-2t}-1}{4t} & t=s, \\
\frac{1-e^{-2t}}{2s} & t<s, \\
\frac{e^{-(t-s)}-e^{-(t+s)}}{2s} & t>s.
\end{cases}
\end{equation}
Note that $q_{\smcp}(t|s)$ is normalized. Curiously, the stationary distribution of the chain is $\pi_{\infty}^{\smcp}(t)=te^{-t}$, exactly as in SMC (Eq. \eqref{eq_stat_pdf}). This can be proved by validating the detailed balance equation, $\pi_{\infty}^{\smcp}(t)q_{\smcp}(s|t)=\pi_{\infty}^{\smcp}(s)q_{\smcp}(t|s)$, which also shows that SMC' is reversible \citep{Zheng2014}.

To define the IBD process (Table \ref{process_smcprime}), we note that in the case $t=s$, the common ancestor of the two chromosomes does not change, and therefore, the shared segment extends until (at least) the next recombination event. 

\begin{algorithm}[H]
\caption{The IBD process under SMC'}
\label{process_smcprime}
\begin{algorithmic}[1]
\State \textbf{Initialize}
\State \hspace{1em} As in Table \ref{process_smc}
\While {$x<L$}
\State $\ell \gets 0$ \Comment \footnotesize The current \emph{total} segment length \normalsize
\Repeat 
\State \hspace{-0.5em} \Comment \footnotesize Draw distance to next recombination; not necessarily a new segment \normalsize
\State Draw $\ell_0 \sim \textrm{Exp}(2Nt)$
\State $\ell \gets \ell + \ell_0$
\State $s\leftarrow t$ 
\State Draw new TMRCA $t$ with PDF $q_{\smcp}(t|s)$ (Eq. \eqref{eq_transition_pdf_smcprime_final})
\Until {$t\ne s$}
\If {$(x+\ell)>L$} \State $\ell \gets (L-x)$ \EndIf
\If {$\ell>m$} 
\State $n_m\gets n_m+1$ 
\State $f_m\gets f_m+\ell/L$ 
\EndIf
\State $x\gets x+\ell$
\EndWhile
\end{algorithmic}
\end{algorithm}

We now derive the stationary distribution of segment lengths. Given the TMRCA $t$ at the beginning of a segment, the rate at which the segment terminates is the product of the recombination rate ($2Nt$) and the probability that the segment does not extend beyond the recombination site ($1-q_{\smcp}(t|t)$). Therefore, given $t$, segment lengths are exponential with rate
\begin{equation}
\lambda(t)=2Nt[1-q_{\smcp}(t|t)]=\frac{N}{2}\left(2t+1-e^{-2t}\right).
\end{equation}
Note that this also implies that for two loci distance $\ell$ apart, and given $t$ at the left locus, the probability of the right TMRCA to remain $t$ is $\exp[-\lambda(t)\ell]=\exp\left\{-\rho t [1-q_{\smcp}(t|t)]\right\}$, as in the small $\rho$ limit of Eq. (30) in \cite{HarrisNielsen}.

To obtain the unconditional distribution of segment lengths, we cannot use $\pi_{\infty}^{\smcp}(t)$, because we need the distribution of tree heights at \emph{segments ends}, not at recombination sites. We therefore define a new Markov chain with transition probability
\begin{equation}
\label{eq_qpp_def}
q_{\smcpseg}(t|s)=\frac{q_{\smcp}(t|s)}{1-q_{\smcp}(s|s)}=\frac{q_{\smcp}(t|s)}{1-\frac{2s+e^{-2s}-1}{4s}},
\end{equation}
which is the \emph{conditional} probability of the new tree height, given that it has changed (i.e., a new segment began). By construction, the stationary distribution of the chain, $\pi_{\infty}^{\smcpseg}(t)$, is the desired distribution of tree heights at the beginning of segments. It is easy to verify by detailed balance that $\pi_{\infty}^{\smcpseg}(t)\propto te^{-t}[1-q_{\smcp}(t|t)]\propto e^{-t}\lambda(t)$, and then, by normalization,
\begin{equation}
\label{eq_pi_stat_smcp_segend}
\pi_{\infty}^{\smcpseg}(t)=\frac{e^{-t}\lambda(t)}{\int_0^{\infty}e^{-t'}\lambda(t')dt'}=\frac{3}{8}e^{-t}\left(2t+1-e^{-2t}\right).
\end{equation}

To obtain the distribution of segment lengths, $\psi_{\smcp}(\ell)$, we integrate over all $t$ (as in Eq. \eqref{eq_seg_len_smc}),
\begin{equation}
\label{eq_psi_smcprime_integral}
\psi_{\smcp}(\ell)=\int_0^{\infty}\pi_{\infty}^{\smcpseg}(t)\lambda(t)e^{-\lambda(t)\ell}dt =
\frac{\int_0^{\infty}\lambda^2(t)e^{-t-\lambda(t)\ell}dt}{\int_0^{\infty}e^{-t}\lambda(t)dt}.
\end{equation}
The integrals in Eq. \eqref{eq_psi_smcprime_integral} can be solved in terms of special functions; the final expression is given in \ref{sect_smcprime_appendix} (Eq. \eqref{eq_smcprime_psi_ell_full}). Note that setting $\lambda(t)=2Nt$ (i.e., setting the probability of $t=s$ to zero) reduces Eq. \eqref{eq_psi_smcprime_integral} to the SMC distribution (Eq. \eqref{eq_seg_len_smc}). Using the representation of Eq. 
\eqref{eq_psi_smcprime_integral}, it is easy to see that $\psi_{\smcp}(\ell)$ is normalized and that the mean segment length is 
\begin{equation}
\label{eq_avgseg_smcprime}
\av{\ell}_{\smcp}=\frac{1}{\int_0^{\infty}e^{-t}\lambda(t)dt}=\frac{3}{4N}.
\end{equation}
Segments in SMC' are, by definition, longer than in SMC, and in SMC, $\psi_{\smc}(\ell)$ had no moments higher than the first. Therefore, $\psi_{\smcp}(\ell)$ also has no second or higher moments. 

It is possible, using Eq. \eqref{eq_psi_smcprime_integral}, to define a renewal process for SMC' analogous to the process defined in Table \ref{process_renewal}. However, with the exception of the infinite-chromosome results (section \ref{sect_ibd_infinite_chr}), we do not further investigate the properties of such a model.

\subsection{Simulations}

\label{sect_intro_simulations}

\begin{figure}
\includegraphics[width=8cm,height=5.5cm]{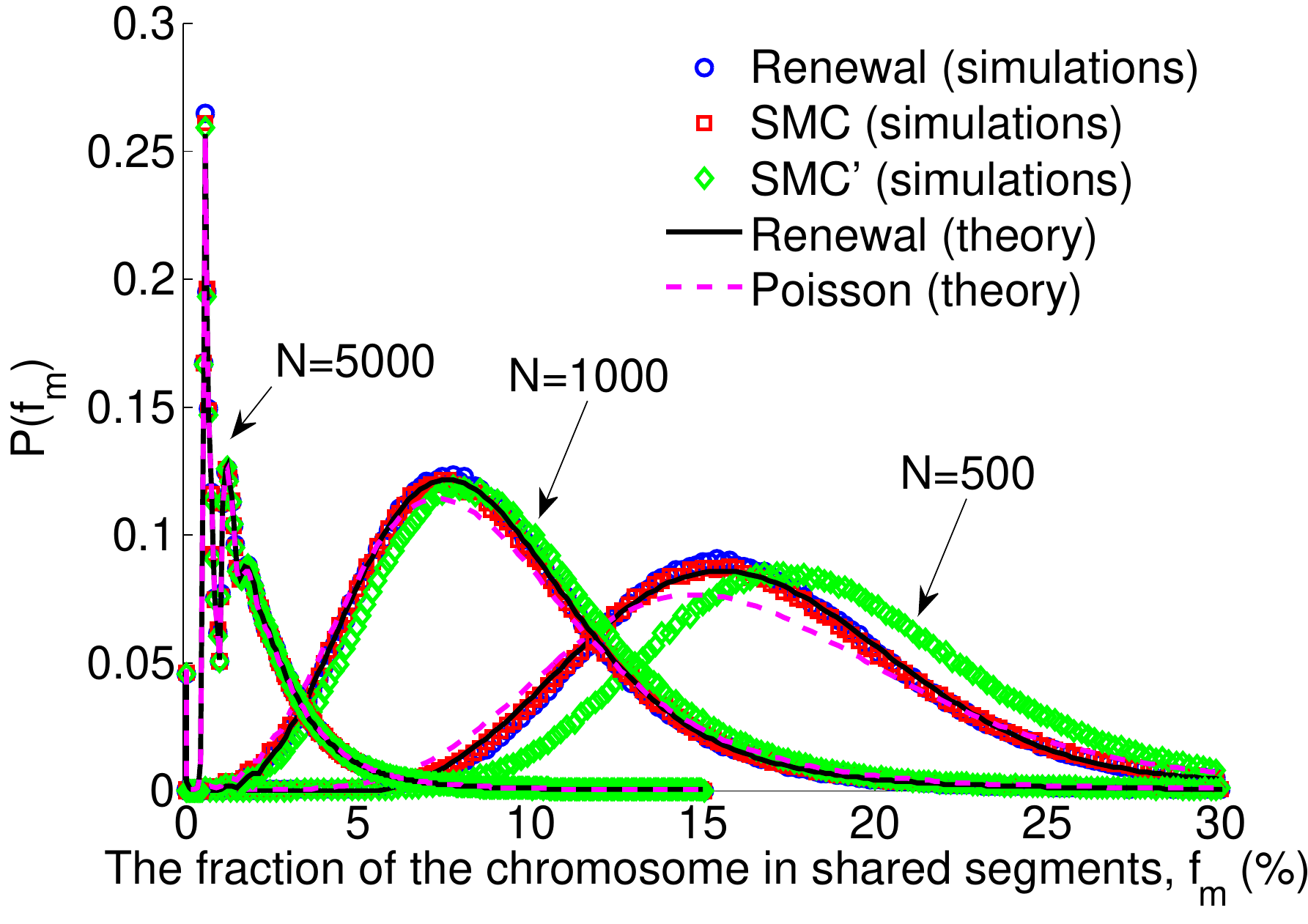}
\caption{The distribution of the fraction of the chromosome found in shared segments longer than $m$, $f_m$. We simulated the IBD process for three values of the population size ($N=500,1000,5000$), for $L=2$ and $m=0.01$ (Morgans), for SMC (the process defined in Table \ref{process_smc}, section \ref{sect_ibd_process_intro}), the renewal approximation (Table \ref{process_renewal}, section \ref{sect_ibd_process_renewal}), and SMC' (Table \ref{process_smcprime}, section \ref{sect_ibd_process_smcprime}), and for $10^6$ realizations for each setting. The distribution for $N=5000$ was divided by 3 for visibility. For all population sizes, SMC and the renewal approximation produced identical results, which also agree well with the renewal theory result (numerical inversion \citep{InverseLaplace2D} of Eq. \eqref{eq_Psu_fracshare_smc}). SMC' and the Poisson approximation (Eq. \eqref{eq_tot_sharing_laplace_poisson}) deviate from SMC/renewal, increasingly for smaller values of $N$. The fluctuations for $N=5000$ are due to the sharing of exactly 0,1,2,... segments of length very close to $m$, and were previously described \citep{HyperSharingGenetics}.} \label{fig_ibd_frac_shared}
\end{figure}

\begin{figure}
\includegraphics[width=8cm,height=6cm]{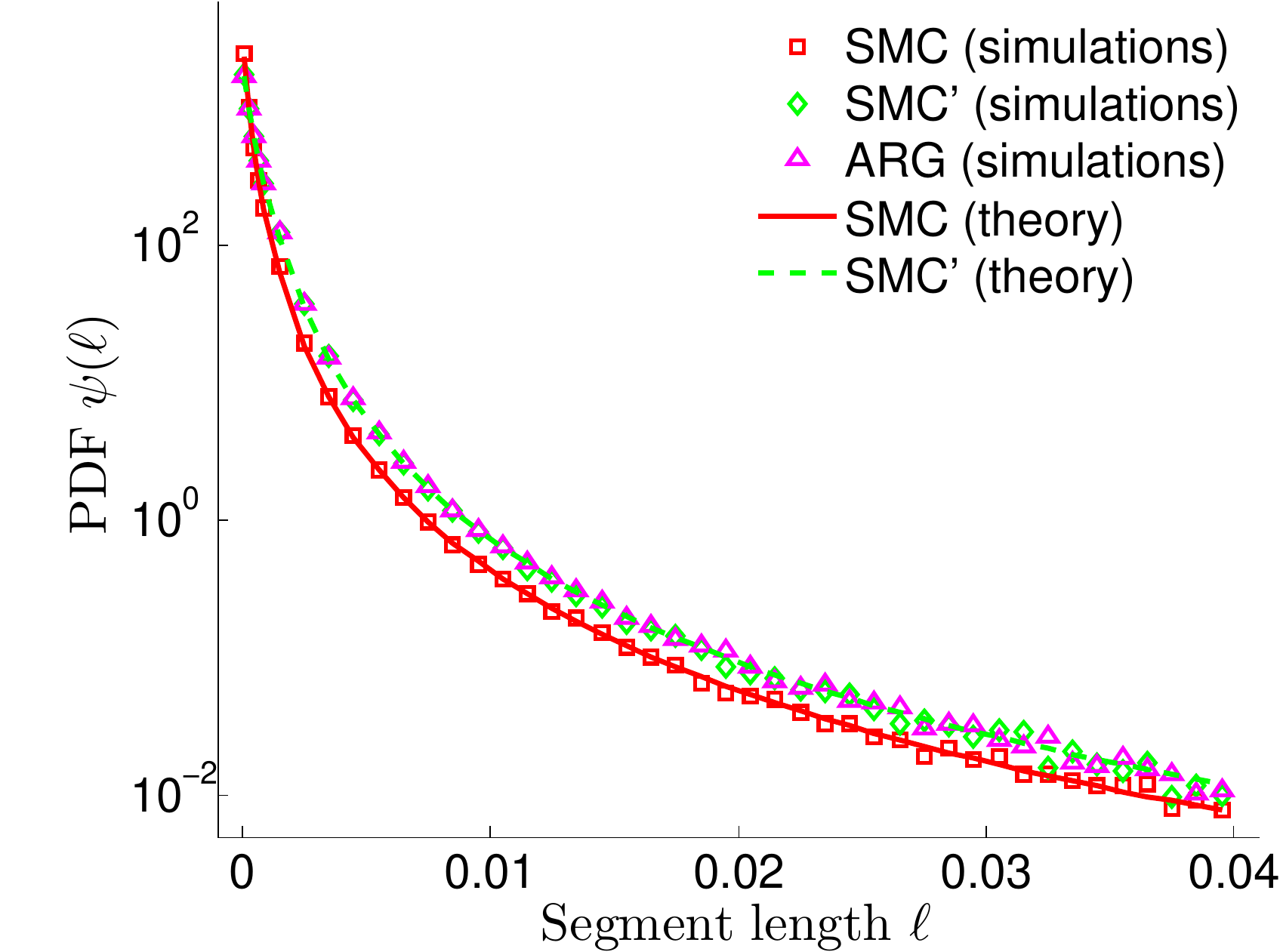}
\caption{The distribution of segment lengths, $\psi(\ell)$, under SMC, SMC', and the ARG. Simulations for SMC and SMC' were as described in Figure \ref{fig_ibd_frac_shared}, but with $N=1000$ and $L=0.5$ (Morgan). ARG simulations were performed in \emph{ms}, by outputting the marginal trees and extracting segment lengths. We ran 5000 realizations for each model. Theory for SMC is from Eq. \eqref{eq_seg_len_smc} and theory for SMC' is from Eq. \eqref{eq_psi_smcprime_integral} (equivalently \eqref{eq_smcprime_psi_ell_full}). Interestingly, simulation results for the ARG are indistinguishable from those of SMC'.} \label{fig_len_pdf}
\end{figure}

To demonstrate the IBD process under SMC and SMC', as well as provide empirical justification to the renewal approximation, we show simulation results for the distribution of the fraction of the chromosome found in shared segments longer than $m$, $P(f_m)$, (Figure \ref{fig_ibd_frac_shared}) and the distribution of segment lengths, $\psi(\ell)$ (Figure \ref{fig_len_pdf}). Simulations were performed precisely as described in Tables \ref{process_smc}, \ref{process_renewal}, and \ref{process_smcprime} above. For all values of $N$ tested, simulation results for $P(f_m)$ were identical between SMC and its renewal approximation. For small values of $N$ (or more precisely, as $1/2N$, the average distance between recombination sites, approaches $m$), there is more sharing in SMC' than in SMC/renewal. This is because in SMC', short segments may extend beyond the first recombination event, and by that exceed the length cutoff. Simulation results for the distribution of segment lengths in SMC and SMC' (Figure \ref{fig_len_pdf}) agree well with Eqs. \eqref{eq_seg_len_smc} and \eqref{eq_psi_smcprime_integral}, respectively. As expected, the SMC' distribution has a heavier tail than in SMC and interestingly, is indistinguishable from that of the ARG, reinforcing the importance of the SMC' model.

\section{The infinite-chromosome limit of the IBD process}

\label{sect_ibd_infinite_chr}

In this section, we derive the mean number of shared segments and the mean fraction of the chromosome in shared segments at the infinite-chromosome limit, under the renewal approximation to  SMC and SMC'. Let us first derive some general, model-independent results. Given a segment length distribution $\psi(\ell)$ and using the elementary renewal theorem (\cite{KarlinTaylor}, Theorem 4.2), the mean total number of segments (of any length) for $L\to \infty$ is 
\begin{equation}
\label{eq_infchr_n0_general}
\av{n_0}=\frac{L}{\av{\ell}}=\frac{L}{\int_0^{\infty}\ell\psi(\ell)d\ell}.
\end{equation}

Using the elementary renewal theorem for reward processes (\cite{KarlinTaylor}, chapter 5, section 7.C.II), the mean number of segments longer than $m$ is, for $L\to \infty$,
\begin{equation}
\label{eq_infchr_nm_general}
\av{n_m}=\av{n_0}\int_m^{\infty}\psi(\ell)d\ell.
\end{equation}
Similarly, the mean fraction of the chromosome found in segments longer than $m$ is
\begin{equation}
\label{eq_infchr_fm_general}
\av{f_m}=\frac{\av{n_0}}{L}\int_m^{\infty}\ell\psi(\ell)d\ell.
\end{equation}
We now turn to specific models, recovering previous results for SMC \citep{Palamara2012} and obtaining new results for SMC'. 

\subsection{The SMC model}

\label{sect_ibd_smc_inf_chr}

Under SMC, the distribution of segment lengths is given by Eq. \eqref{eq_seg_len_smc}.
The mean total number of segments is
\begin{equation}
\label{eq_n0_avg_smc}
\av{n_0}_{\smc}=\frac{L}{\int_0^{\infty}\frac{4N\ell}{(1+2N\ell)^3}d\ell}=2NL.
\end{equation}
The mean number of shared segments longer than $m$ is
\begin{equation}
\label{eq_ibd_avg_segnum}
\av{n_m}_{\smc}=2NL\int_m^{\infty}\frac{4N}{(1+2N\ell)^3}d\ell=\frac{2NL}{(1+2mN)^2}.
\end{equation}
The mean fraction of the chromosome in segments longer than $m$ is
\begin{equation}
\label{eq_avg_sharing}
\av{f_m}_{\smc}=2N\int_m^\infty \frac{4N\ell}{(1+2N\ell)^3}d\ell=\frac{1+4mN}{(1+2mN)^2}.
\end{equation}

\begin{figure}
\includegraphics[width=7cm,height=5.5cm]{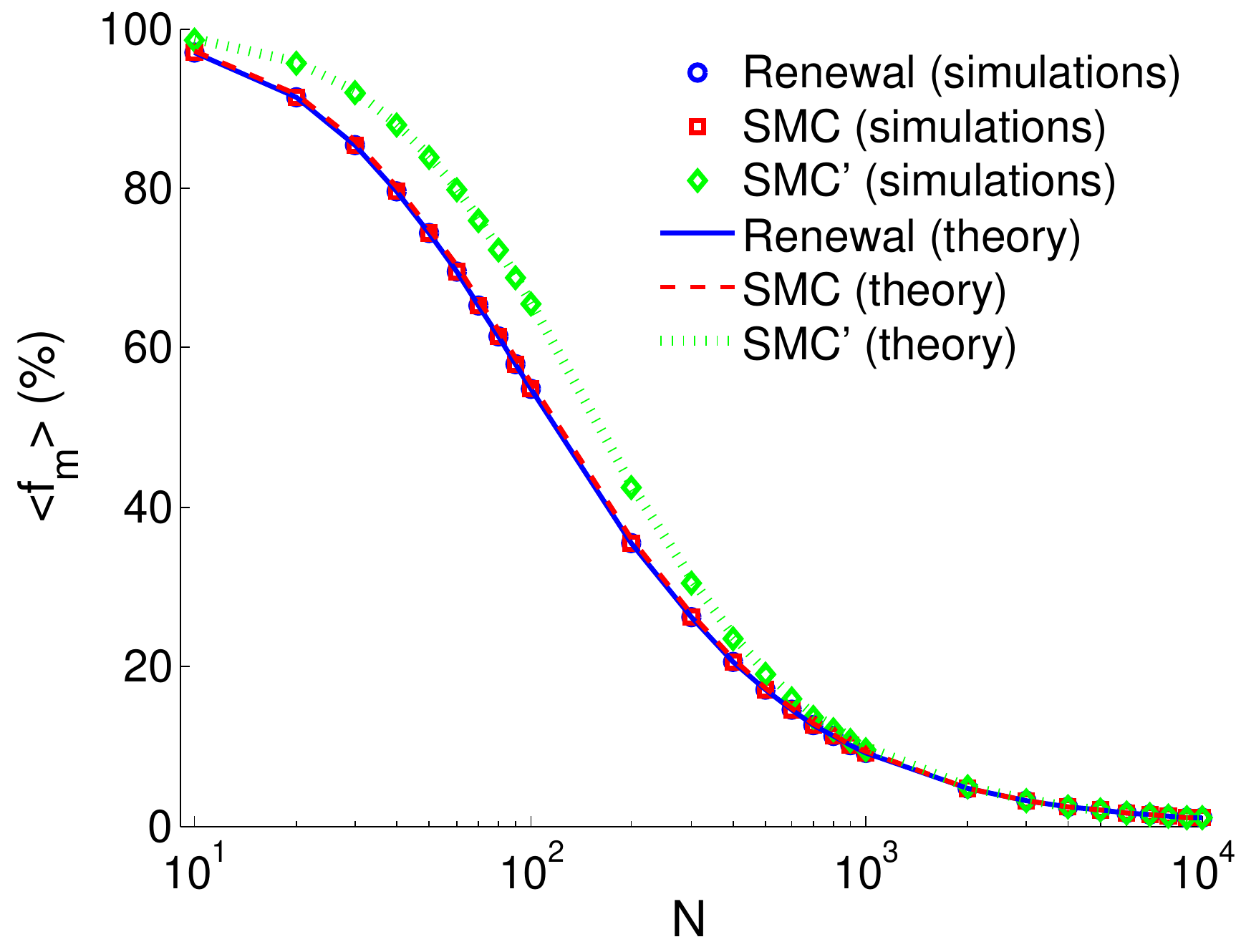}
\caption{The mean fraction of the chromosome found in shared segments longer than $m$, $\av{f_m}$. Simulation details are as in Figure \ref{fig_ibd_frac_shared}. Simulation results and theory for SMC and the renewal approximation coincide. The renewal theory curve was obtained by numerically inverting \citep{deHoog} Eq. \eqref{eq_avg_fracshare_s_renewal}. Theory for SMC and SMC' (infinite-chromosome limits) is from Eqs. \eqref{eq_avg_sharing} and \eqref{eq_avg_sharing_smcprime}, respectively.} \label{fig_ibd_avg_sharing}
\end{figure}

Eq. \eqref{eq_avg_sharing} has been previously derived by \cite{Palamara2012}, by studying the distribution of segment lengths surrounding a randomly chosen site. Simulation results for $\av{f_m}_{\smc}$ (Figure \ref{fig_ibd_avg_sharing}) agree well with Eq. \eqref{eq_avg_sharing}. While simulations were shown before \citep{Palamara2012,HyperSharingGenetics}, here we are able to observe perfect agreement even for very small values of $N$. Eq. \eqref{eq_ibd_avg_segnum} was derived by \cite{Palamara2012} using the relation $\av{n_m}=L\av{f_m}/\av{\ell_m}$, where $\av{\ell_m}$ is the mean length of segments longer than $m$.

Eq. \eqref{eq_ibd_avg_segnum} can be derived in yet another way, using a result from \cite{RalphCoop}, who showed that for a fixed TMRCA $t$, the mean number of segments longer than $m$ is $K(t,m)=e^{-2mNt}[2Nt(L-m)+1]$. Integrating over all $t$ using $P_c(t)=e^{-t}$, we have $\av{n_m}=\int_0^{\infty}K(t,m)P_c(t)dt=(1+2NL)/(1+2mN)^2$. For $L\gg 1/2N$, we recover Eq. \eqref{eq_ibd_avg_segnum}. Also note that for a fixed $t$, the mean number of segments of length in $[\ell,\ell+d\ell]$ is $-\di K(t,\ell)/\di \ell\,d\ell$. Integrating over all $t$ as before, this gives $4N(1+2NL)/(1+2N\ell)^3\,d\ell$. Since the total number of segments (of all lengths) is $K(t,0)=(1+2NL)$, the probability of a random segment to be of length in $[\ell,\ell+d\ell]$ is $\psi(\ell)d\ell=4N/(1+2N\ell)^3\,d\ell$, exactly as in our Eq. \eqref{eq_seg_len_smc}.

\subsection{The SMC' model}

\label{sect_ibd_smcprime_avg_sharing}

Under SMC', the distribution of segment lengths is given by Eq. \eqref{eq_psi_smcprime_integral}. The mean total number of segments is (using Eq. \eqref{eq_avgseg_smcprime})
\begin{equation}
\label{eq_numseg_avg_smcprime}
\av{n_{0}}_{\smcp}=\frac{L}{\int_0^{\infty}\ell\psi_{\smcp}(\ell)d\ell}=\frac{4NL}{3}.
\end{equation}
Eq. \eqref{eq_numseg_avg_smcprime} represents a surprisingly simple result, stating that for long chromosomes, the mean number of segments in SMC' is  precisely $2/3$ of the total number of recombination events ($2NL$). To provide an intuitive explanation, we recall (section \ref{sect_ibd_process_smcprime}) that the stationary distribution of tree heights at recombination sites in SMC' is $\pi_{\infty}^{\smcp}(t)=te^{-t}$ (as in SMC). At a recombination site, there is probability $1-q_{\smcp}(t|t)$ for the TMRCA to change and consequently, for the segment to terminate. Integrating over all $t$,
\begin{align}
&\int_0^{\infty}te^{-t}[1-q_{\smcp}(t|t)]dt=\nonumber \\ &
\int_0^{\infty}te^{-t}\frac{2t+1-e^{-2t}}{4t}dt= \frac{2}{3}.
\end{align}
In fact, it can be shown that at stationarity, the new tree has equal probability to be either larger, smaller, or equal to the previous tree. Also note that the probability to change the MRCA at a recombination site is $2/3$ also for the ARG (\cite{GriffithsMarjoram1997}, Theorem 2.4).

Next, using Eqs. \eqref{eq_psi_smcprime_integral}, \eqref{eq_avgseg_smcprime}, and \eqref{eq_infchr_n0_general}, it can be seen that 
\begin{equation}
\label{eq_psiell_smcprime_integral_n0}
\psi_{\smcp}(\ell)=\frac{\int_0^{\infty}\lambda^2(t)e^{-t-\lambda(t)\ell}dt}{\av{n_{0}}_{\smcp}/L}.
\end{equation}
Using Eqs. \eqref{eq_infchr_nm_general} and \eqref{eq_psiell_smcprime_integral_n0}, the mean number of segments longer than $m$ is
\begin{equation}
\label{eq_nm_smcprime_integral}
\av{n_{m}}_{\smcp}=\av{n_{0}}_{\smcp}\int_m^{\infty}\psi_{\smcp}(\ell)d\ell=L\int_0^{\infty}\lambda(t)e^{-t-\lambda(t)m}dt.
\end{equation}
The final result, which we obtained using \textsc{Mathematica} \citep{Mathematica}, is given in \ref{sect_smcprime_appendix} (Eq. \eqref{eq_smcprime_avg_nm_full}).

Finally, using Eqs. \eqref{eq_infchr_fm_general} and \eqref{eq_psiell_smcprime_integral_n0}, we have 
\begin{equation}
\label{eq_avg_sharing_smcprime_integral}
\av{f_{m}}_{\smcp}=\frac{\av{n_{0}}_{\smcp}}{L}\int_m^{\infty}\ell\psi_{\smcp}(\ell)d\ell=\int_0^{\infty}e^{-t-\lambda(t)m}[1+\lambda(t) m]dt.
\end{equation}
The result of the integral is given in \ref{sect_smcprime_appendix} (Eq. \eqref{eq_avg_sharing_smcprime}). Numerical evaluation shows perfect agreement with simulation results, for all values of $N$ (Figure \ref{fig_ibd_avg_sharing}).

\section{Renewal theory results for finite chromosomes}

\label{sect_finite_chr}

In this section, we use renewal theory to derive the complete distribution of our quantities of interest: the number of segments longer than $m$ (section \ref{sect_ibd_numseg_pdf}) and the fraction of the chromosome in segments longer than $m$ (section \ref{sect_ibd_fracshared_pdf}), for a chromosome of a finite size $L$. In both cases, we derive an expression in Laplace space for the distribution (Eq. \eqref{eq_Pn_s_psi_final} for the number of segments and Eq. \eqref{eq_Psu_fracshare_laplace} for the fraction of the chromosome). Those expressions are general for any segment length distribution. We then substitute the specific SMC form, to obtain explicit expressions (\ref{sect_renewal_appendix}). As we show, the distributions can be numerically inverted and compared to simulations or be used for demographic inference. Using standard techniques, we also obtain the first two moments (in real space) for long (but finite) chromosomes. Our method in this section is adapted from the physics literature \citep{GodrecheLuck}.

\subsection{The distribution of the number of segments longer than $m$ under the renewal approximation}

\label{sect_ibd_numseg_pdf}

\subsubsection{Theory}

\label{sect_ibd_numseg_dist_derive}

Define $P(n_m=k;L)$ as the probability that two chromosomes share exactly $k$ segments longer than $m$ over a sequence of length $L$, under the renewal IBD process defined in Table \ref{process_renewal} (section \ref{sect_ibd_process_renewal}). We will obtain $\tP(n_m=k,s)$, the Laplace transform of $P(n_m=k,L)$ with respect to $L$: $\tP(n_m=k,s)=\int_0^{\infty}e^{-sL}P(n_m=k,L)dL$. Let us first define an auxiliary function, $\eta_m(L)dL$, which is the probability that, conditional on recombination at position 0 in the sequence, \begin{inparaenum}[\itshape a\upshape)]
\item recombination occurred at position in $[L,L+dL]$; and
\item all intermediate recombination events in $[0,L]$ had terminated segments that were shorter than $m$.
\end{inparaenum} Note that $\eta_m(L)$, as well as $Q_m(k,L)$ below (Eq. \eqref{eq_Qn_rec_real}), are not PDFs. Then, $\eta_m(L)$ satisfies
\begin{equation}
\label{eq_eta_L_real}
\eta_m(L)=\delta(L)+\int_0^{\min(m,L)} \psi(\ell)\eta_m(L-\ell)d\ell.
\end{equation}
In Eq. \eqref{eq_eta_L_real}, $\delta(x)$ is Dirac's delta function and $\psi(\ell)$ is the PDF of segment lengths. The derivation will proceed with a general $\psi(\ell)$; we will substitute the explicit SMC form (Eq. \eqref{eq_seg_len_smc}) only at the final result. Eq. \eqref{eq_eta_L_real} is explained as follows. The first term ($\delta(L)$) accounts for the case $L=0$. Otherwise, we condition on the length of the last segment, $\ell$, which cannot exceed either $m$ or $L$. Given $\ell$, we require the recombination at $L-\ell$ to end a series of short segments, which happens with probability $\eta_m(L-\ell)$. Note that we made use of the renewal property, namely the independence of successive segment lengths.

We now apply the Laplace transform ($L\to s$) to both sides of Eq. \eqref{eq_eta_L_real},
\begin{align}
\label{eq_eta_s_derive}
\teta_m(s) &= 1 + \int_0^{\infty}e^{-sL}\left[\int_0^{\min(m,L)} \psi(\ell)\eta_m(L-\ell)d\ell\right]dL \nonumber \\
&= 1+\int_0^m\left[\int_{\ell}^{\infty}e^{-sL}\psi(\ell)\eta_m(L-\ell)dL \right]d\ell \nonumber\\
&= 1+\int_0^m e^{-s\ell}\psi(\ell)\left[\int_{\ell}^{\infty}e^{-s(L-\ell)}\eta_m(L-\ell)dL \right]d\ell \nonumber\\
&= 1+\int_0^m e^{-s\ell}\psi(\ell)d\ell \int_{0}^{\infty}e^{-sL'}\eta_m(L')dL' \nonumber\\
&=1+\tpsi_{<m}(s)\teta_m(s),
\end{align}
where we defined $\tpsi_{<m}(s)\equiv \int_0^m e^{-s\ell}\psi(\ell)d\ell$. We thus obtained an algebraic equation for $\teta_m(s)$, whose solution is
\begin{equation}
\label{eq_eta_s_final}
\teta_m(s)=\left[1-\tpsi_{<m}(s)\right]^{-1}.
\end{equation}

Next, we define another auxiliary function, $Q_m(k,L)dL$, which is the probability that \begin{inparaenum}[\itshape a\upshape)]
\item recombination occurred at position in $[L,L+dL]$; and
\item that the recombination event of (a) has ended the $k$th segment \emph{longer than $m$}. \end{inparaenum}
For $k=0$, ~$Q_m(0,L)=\delta(L)$. For $k>0$, we have the following recursion equation,
\begin{equation}
\label{eq_Qn_rec_real}
Q_m(k,L)= \int_m^L\psi(\ell)\left[\int_0^{L-\ell}\eta_m(\ell')Q_m(k-1,L-\ell-\ell')d\ell'\right]d\ell.
\end{equation}
Eq. \eqref{eq_Qn_rec_real} is explained similarly to Eq. \eqref{eq_eta_L_real}. We condition on the length of the last segment, $\ell$, which must be longer than $m$ (but shorter than $L$). Given the preceding recombination at $L-\ell$, we condition on the length of rightmost stretch of short segments, $\ell'$, which has probability $\eta_m(\ell')$. Note that $\eta_m(L)$ does not depend on the absolute position along the sequence, again, due to the renewal property. Finally, given $\ell$ and $\ell'$, there must have been a recombination event at $L-\ell-\ell'$ ending the $(k-1)$th segment longer than $m$, with probability $Q_m(k-1,L-\ell-\ell')$. We now apply the Laplace transform to Eq. \eqref{eq_Qn_rec_real},
\begin{align}
\label{eq_Q_s_derive}
\tQ_m(k,s)&=\int_m^{\infty}e^{-sL} \left\{\int_m^L\psi(\ell)\left[\int_0^{L-\ell}\eta_m(\ell')Q_m(k-1,L-\ell-\ell')d\ell'\right]d\ell\right\}dL \nonumber \\ &=
\int_m^{\infty}e^{-s\ell}\psi(\ell)d\ell \int_{\ell}^{\infty}e^{-s(L-\ell)}\left[\int_0^{L-\ell}\eta_m(\ell')Q_m(k-1,L-\ell-\ell')d\ell'\right]dL
\nonumber \\ &=
\int_m^{\infty}e^{-s\ell}\psi(\ell)d\ell \int_{0}^{\infty}e^{-sL'}\left[\int_0^{L'}\eta_m(\ell')Q_m(k-1,L'-\ell')d\ell'\right]dL' \nonumber \\ &=
\tpsi_{>m}(s)\teta_m(s)\tQ_m(k-1,s)=\frac{\tpsi_{>m}(s)}{1-\tpsi_{<m}(s)}\tQ_m(k-1,s),
\end{align}
where $\tpsi_{>m}(s)\equiv \int_m^{\infty} e^{-s\ell}\psi(\ell)d\ell$, we used the fact that $Q_m(k>0,L<m)=0$, and in the last line, we used the convolution theorem and Eq. \eqref{eq_eta_s_final}. Using Eq. \eqref{eq_Q_s_derive} and the initial condition, $\tQ_m(0,s)=1$, we have
\begin{equation}
\label{eq_Qn_s_final}
\tQ_m(k,s)=\left(\frac{\tpsi_{>m}(s)}{1-\tpsi_{<m}(s)}\right)^k.
\end{equation}
We next define $\phi(\ell)\equiv 1-\int_0^{\ell}\psi(\ell')d\ell'=\int_{\ell}^{\infty}\psi(\ell')d\ell'$, the probability that a segment extends for sequence length greater than $\ell$. We are now in a position to compute $P(n_m=k,L)$. For $k>0$,
\begin{align}
\label{eq_Pn_real}
P(n_m=k,L)=& \int_0^{m}\phi(\ell)\left[\int_0^{L-\ell}\eta_m(\ell')Q_m(k,L-\ell-\ell')d\ell'\right]d\ell \nonumber \\ &+
\int_m^{L}\phi(\ell)\left[\int_0^{L-\ell}\eta_m(\ell')Q_m(k-1,L-\ell-\ell')d\ell'\right]d\ell.
\end{align}
For $P(n_m=k,L)$, we do not require recombination at $L$. Therefore, we condition on the sequence length $\ell$ since the rightmost recombination event, with the probability of no recombination since then being $\phi(\ell)$. Then, if $\ell<m$, we require $k$ segments longer than $m$ to be seen by position $L-\ell$, possibly followed by any number of short segments. If $\ell>m$, then the sequence $[L-\ell,L]$ will form a segment longer than $m$ on its own, and we only require $k-1$ previous segments longer than $m$. Eq. \eqref{eq_Pn_real} can be transformed similarly to Eqs. \eqref{eq_eta_s_derive} and \eqref{eq_Q_s_derive}, yielding
\begin{equation}
\label{eq_Pn_s_final}
\tP(n_m=k,s)=\teta_m(s)\left[\tphi_{<m}(s)\tQ_m(k,s)+\tphi_{>m}(s)\tQ_m(k-1,s)\right],
\end{equation}
where $\tphi_{<m}(s)=\int_0^me^{-s\ell}\phi(\ell)d\ell$ and $\tphi_{>m}(s)=\int_m^{\infty}e^{-s\ell}\phi(\ell)d\ell$. For $k=0$, we have $P(n_m=0,L)=\int_0^{\min(m,L)}\phi(\ell)\eta_m(L-\ell)d\ell$. Applying the Laplace transform gives 
\begin{equation}
\label{eq_P0_s_final}
\tP(n_m=0,s)=\tphi_{<m}(s)\teta_m(s).
\end{equation}
Combining Eqs. \eqref{eq_eta_s_final}, \eqref{eq_Qn_s_final}, \eqref{eq_Pn_s_final}, and \eqref{eq_P0_s_final}, and using $\tphi_{<m}(s)+\tphi_{>m}(s)=\tphi(s)=[1-\tpsi(s)]/s$ and $\tpsi_{<m}(s)+\tpsi_{>m}(s)=\tpsi(s)$, we finally obtain
\begin{equation}
\label{eq_Pn_s_psi_final}
\tP(n_m=k,s)=\begin{cases}
\frac{\tphi_{<m}(s)}{1-\tpsi_{<m}(s)} & k=0, \\\frac{[1-\tpsi(s)][\tpsi_{>m}(s)+s\tphi_{>m}(s)]}{s[1-\tpsi_{<m}(s)]^2}\left[\frac{\tpsi_{>m}(s)}{1-\tpsi_{<m}(s)}\right]^{k-1} & k>0.
\end{cases}
\end{equation}
Eq. \eqref{eq_Pn_s_psi_final} is our main result, and is valid for any distribution of segment lengths, $\psi(\ell)$. 
Due to normalization, we expect $\sum_{k=0}^{\infty}\tP(n_m=k,s)=\sum_{k=0}^{\infty}\int_0^{\infty}e^{-sL}P(n_m=k,L)dL=\int_0^{\infty}e^{-sL}\left[\sum_{k=0}^{\infty}P(n_m=k,L)\right]dL=\int_0^{\infty}e^{-sL}dL=1/s$, as can be verified, after some algebra, from Eq. \eqref{eq_Pn_s_psi_final}.

Our results have so far been general and could apply to any `IBD process'. We now substitute the SMC segment length PDF, $\psi(\ell)=4N/(1+2N\ell)^3$ (Eq. \eqref{eq_seg_len_smc}). The distribution of the number of segments longer than $m$ (Eq. \eqref{eq_Pn_s_psi_final}) under SMC is given in Eq. \eqref{eq_Pn_s_smc} (\ref{sect_renewal_appendix}). This can be numerically inverted \citep{deHoog}, for each $k$, to obtain, for a given $L$, the distribution $P(n_m=k)$. The theoretical prediction compares perfectly to simulation results for both SMC and the renewal approximation (Figure \ref{fig_ibd_numseg_dist}).

\subsubsection{The mean}

\label{sect_numseg_mean}

The mean number of segments longer than $m$ is $\av{n_m}=\sum_{k=0}^{\infty}kP(n_m=k,L)$. Taking the Laplace transform of $\av{n_m}$, using Eq. \eqref{eq_Pn_s_psi_final} and the relation $\sum_{k=0}^{\infty}kx^k=x/(1-x)^2$, we obtain, after some algebra,
\begin{equation}
\label{eq_numseg_avg_s}
\tilde{\av{n_m}}(s)=\frac{\tpsi_{>m}(s)+s\tphi_{>m}(s)}{s[1-\tpsi(s)]}.
\end{equation}

For SMC, we obtain, using Eq. \eqref{eq_numseg_avg_s} and \textsc{Mathematica},
\begin{equation}
\tilde{\av{n_m}}_{\smc}(s)=\frac{4N^2e^{-ms}}{s^2(1+2mN)^2\left[se^{\frac{s}{2N}}\text{Ei}\left(-\frac{s}{2N}\right)+2N\right]},
\end{equation}
where $\text{Ei}$ is the exponential integral function. Noting that $\lim_{s\to 0}s^2\tilde{\av{n_m}}_{\smc}(s)=2N/(1+2mN)^2$, we have $\lim_{L\to \infty}\av{n_m}_{\smc}/L=2N/(1+2mN)^2$, exactly as in Eq. \eqref{eq_ibd_avg_segnum}. 

\subsubsection{The variance}

\label{sect_numseg_var}

The second moment of the number of segments longer than $m$ can be computed using $\tilde{\av{n_m^2}}(s)=\sum_{k=0}^{\infty}k^2\tP(n_m=k,s)$, from which the variance can be obtained. For SMC and for large $L$,
\begin{align}
\label{eq_numseg_var}
\Var[n_m]_{\smc}&=\frac{2NL}{(1+2mN)^4}\left[2\ln(2NL)+4mN(mN-1)-5\right] +O(\ln^2 L).
\end{align}


\begin{figure}
\includegraphics[width=8cm,height=5.5cm]{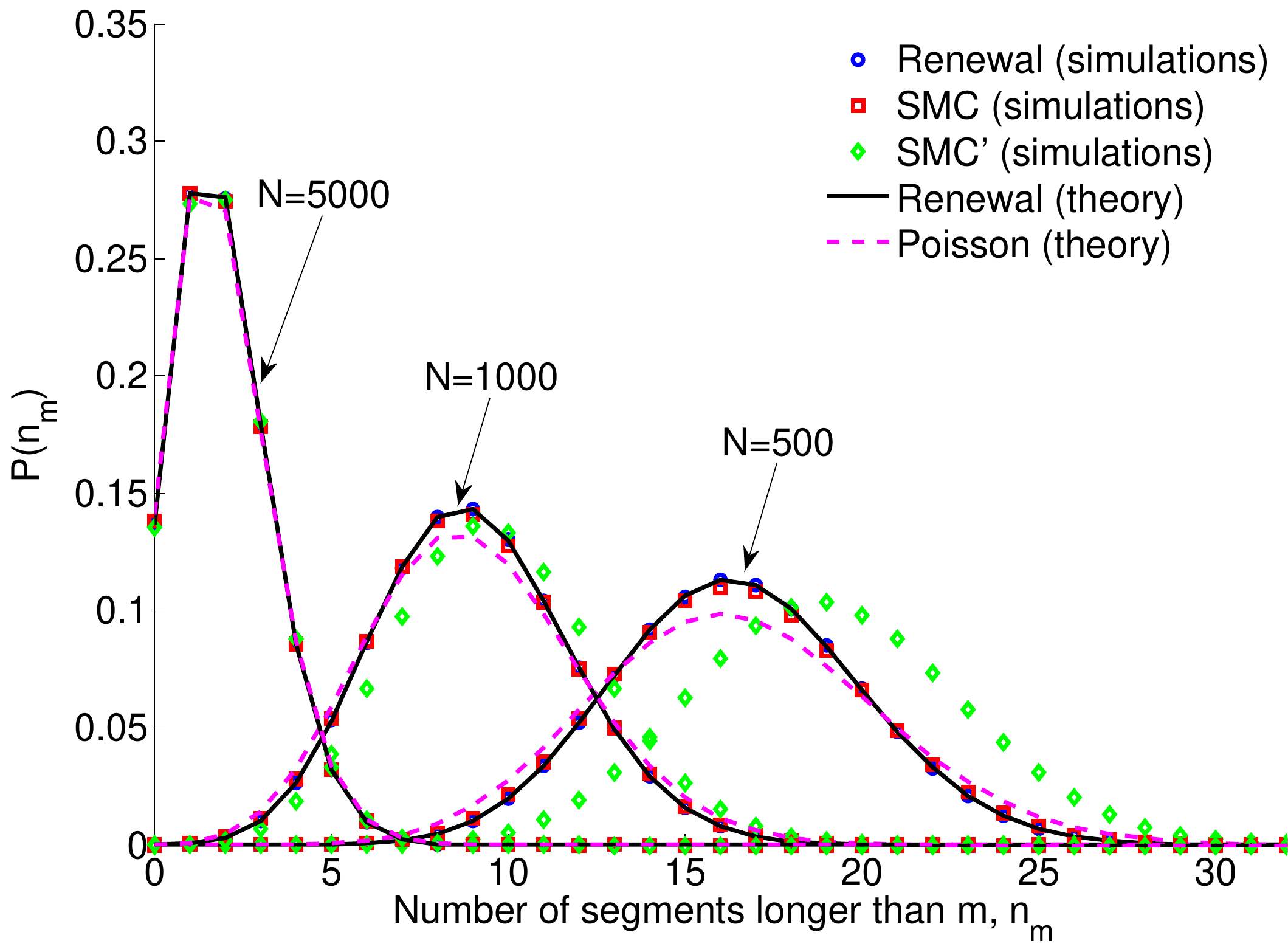}
\caption{The distribution of the number of shared segments longer than $m$, $n_m$. Simulation details are as in Figure \ref{fig_ibd_frac_shared} (specifically, $L=2$ and $m=0.01$ (Morgans)). Theory for the renewal approximation was obtained by numerically inverting Eq. \eqref{eq_Pn_s_smc}. The Poisson distribution has mean $2NL/(1+2mN)^2$ (Eq. \eqref{eq_ibd_avg_segnum}).} \label{fig_ibd_numseg_dist}
\end{figure}

\subsubsection{The Poisson approximation}

\label{sect_numseg_poisson}

\cite{Palamara2012}, following \cite{Huff_ERSA}, proposed that the number of shared segments longer than $m$ is Poisson distributed, with the infinite-chromosome mean, $\av{n_m}_{\smc}=2NL/(1+2mN)^2$ (Eq. \eqref{eq_ibd_avg_segnum}). The Poisson distribution fits the simulation results reasonably (Figure \ref{fig_ibd_numseg_dist}; see also section \ref{sect_fracshare_poisson}). Indeed, for large values of $N$ and $L$, Eq. \eqref{eq_numseg_var} gives $\Var[n_m]_{\smc}\approx \av{n_m}_{\smc}\approx\frac{L}{2m^2N}$, as expected from a Poisson variable. Deviations appear for small values of $N$ (Figure \ref{fig_ibd_numseg_dist}). 

\subsubsection{Demographic inference}

\label{sect_ibd_inference}

\begin{figure}
\includegraphics[width=7cm,height=5.5cm]{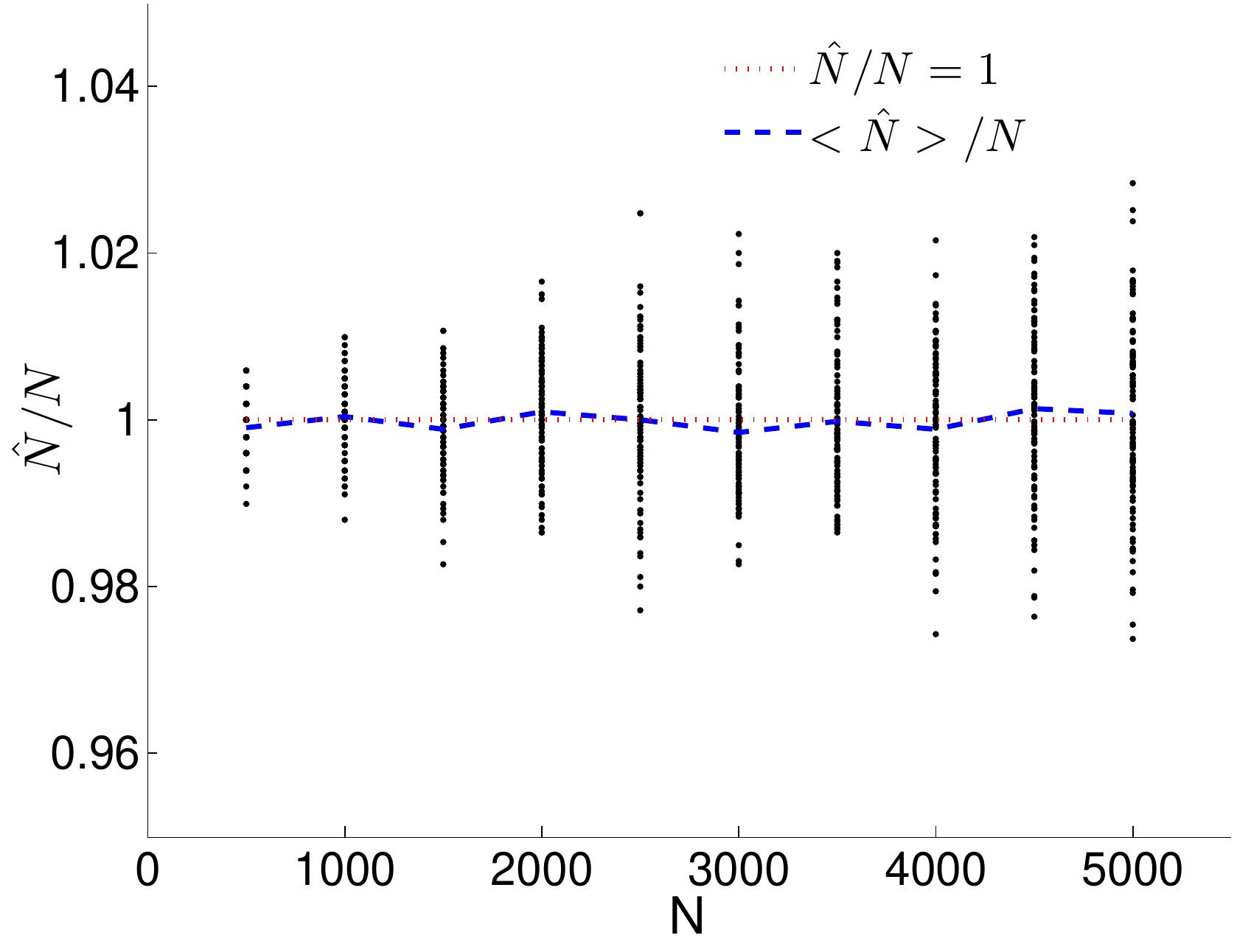}
\caption{Inference of the effective population size using the distribution of the number of shared segments longer than $m$. Simulations for $N=500,1000,...,5000$ were performed as in Figure \ref{fig_ibd_frac_shared} and for $R=5000$ pairs of chromosomes, and Eq. \eqref{eq_mle_N} was used to compute $\hat{N}$, the estimator of the population size. We then repeated 100 times for each $N$, and each ratio $\hat{N}/N$ is shown as a dot. The dotted red line represents $\hat{N}=N$ and the blue line shows $\av{\hat{N}}/N$. The estimator in unbiased, with standard deviation as low as $0.003N$ for $N=500$ and $0.011N$ for $N=5000$.} \label{fig_ibd_inference}
\end{figure}

The results of section \ref{sect_ibd_numseg_dist_derive} have attractive implications for demographic inference. While this is not our main focus here, we provide a simple demonstration. For a given population size $N$ (and for $L=2$ and $m=0.01$ (Morgans)), we simulated the SMC IBD process $R=5000$ times and recorded, for each run, the number of shared segments longer than $m$, $n_m$. This corresponds, roughly, to the information that will be available from sampling a single chromosome in $50$ (diploid) individuals, although we note that in reality, pairs of chromosomes in a sample are weakly dependent (see \cite{HyperSharingGenetics} and the Discussion). Additionally, the underlying ancestral process is neither SMC nor even the coalescent with recombination, but there is rather a shared underlying pedigree \citep{Wakeley_pedigree}; however, we leave investigation of more complex models to future studies. Given $N$, $m$, and $L$, the log-likelihood of the sample $\{n^{(i)}_m\}$, $i=1,...,R$, is
\begin{equation}
\textrm{LL}(N)=\sum_{i=1}^{R}\log P\left(n_m=n^{(i)}_m,L\right),
\end{equation}
where $P(n_m=k,L)$ is given by numerically inverting, $s\to L$, Eq. \eqref{eq_Pn_s_smc}. We then computed the \emph{maximum likelihood estimator},
\begin{equation}
\label{eq_mle_N}
\hat{N}=\underset{N}{\arg\max} ~\textrm{LL}(N).
\end{equation}
Simulation results (Figure \ref{fig_ibd_inference}) show that the estimator performs excellently, with standard deviation $\approx 0.01N$ or lower. The performance of the estimator deteriorates for large values of $N$, since the number of shared segments longer than $m$ approaches zero (Figure \ref{fig_ibd_numseg_dist}; Eq. \eqref{eq_ibd_avg_segnum}). Under our ``noise-free'' simulations, even the simple-minded estimator, $\hat{N}=1/\left(m\av{f_m}\right)-3/(4m)$ \citep{HyperSharingGenetics}, performs well, although with bias ($\av{\hat{N}}/N\approx 1.02$; see \cite{HyperSharingGenetics}) and with $\approx 60\%$ larger standard deviation than the maximum likelihood estimator.

\subsection{The distribution of the fraction of the chromosome found in segments longer than $m$}

\label{sect_ibd_fracshared_pdf}

\subsubsection{Theory}

\label{sect_ibd_frac_shared_derive}

Denote $P(f_m)$ as the density of the fraction of the chromosome found in shared segments longer than $m$. The derivation of $P(f_m)$ uses techniques similar to those used in section \ref{sect_ibd_numseg_dist_derive} and is tedious. We therefore omit the details and skip to the analysis of the final result. Let $P(L_m,L)$ be the density of $L_m\equiv Lf_m$, the total sequence length found in shared segments longer than $m$, given a chromosome of length $L$, and let $\tP_{L_m}(u,s)$ be its Laplace transform. This is a double Laplace transform: $L\to s$ and $L_m\to u$, or $\tP_{L_m}(u,s)=\int_0^{\infty}\int_0^{\infty}e^{-uL_m-sL}P(L_m,L)dL_mdL$. For the renewal IBD process defined in section \ref{sect_ibd_process_renewal} and with segment length PDF $\psi(\ell)$, it can be shown that
\begin{equation}
\label{eq_Psu_fracshare_laplace}
\tP_{L_m}(u,s)=\frac{\frac{1}{s}-\frac{1}{s}\tpsi_{<m}(s)+\phi(m)\left[\frac{e^{-m(s+u)}}{s+u}-\frac{e^{-ms}}{s}\right]-\frac{\tpsi_{>m}(s+u)}{s+u}}{1-\tpsi_{<m}(s)-\tpsi_{>m}(s+u)},
\end{equation}
where, as in section \ref{sect_ibd_numseg_dist_derive}, $\phi(\ell)=1-\int_0^{\ell}\psi(\ell')d\ell'$, $\tpsi_{<m}(z)=\int_0^me^{-z\ell}\psi(\ell)d\ell$, and $\tpsi_{>m}(z)=\int_m^{\infty}e^{-z\ell}\psi(\ell)d\ell$. For $u=0$, we expect, due to normalization, $\tP_{L_m}(u=0,s)=\int_0^{\infty}e^{-sL}\int_0^{\infty}P(L_m,L)dL_mdL=\int_0^{\infty}e^{-sL}dL=1/s$, as can be verified from Eq. \eqref{eq_Psu_fracshare_laplace}. 

We then substituted the SMC form, $\psi(\ell)=4N/(1+2N\ell)^3$ (Eq. \eqref{eq_seg_len_smc}), and evaluated Eq. \eqref{eq_Psu_fracshare_laplace} in \textsc{Mathematica}. The final result is given in \ref{sect_renewal_appendix}, Eq. \eqref{eq_Psu_fracshare_smc}.
Eq. \eqref{eq_Psu_fracshare_smc} can be numerically inverted with respect to both $u$ and $s$ \citep{InverseLaplace2D} to give $P(L_m,L)$, from which we have $P(f_m=L_m/L)=LP(L_m,L)$. The theoretical prediction agrees well with simulations (Figure \ref{fig_ibd_frac_shared}). Very small deviations may be due to numerical errors in the two-dimensional inversion.

\subsubsection{The mean}

\label{sect_fracshared_mean}

The mean sequence length in segments longer than $m$, $\av{L_m}$, can be obtained (in $s$ space) from $\tP_{L_m}(u,s)$ by
\begin{equation}
\tilde{\av{L_m}}(s)=-\left.\frac{\di \tP_{L_m}(u,s)}{\di u}\right|_{u=0}.
\end{equation}
For a general $\psi(\ell)$, we obtain from Eq. \eqref{eq_Psu_fracshare_laplace},
\begin{equation}
\tilde{\av{L_m}}(s)=\frac{\phi(m)e^{-ms}(1+ms)-\tpsi_{>m}(s)}{s^2\left[1-\tpsi(s)\right]}.
\end{equation}
For the SMC form of $\psi(\ell)$ (Eq. \eqref{eq_seg_len_smc}), this gives
\begin{equation}
\label{eq_avg_fracshare_s_renewal}
\tilde{\av{L_m}}_{\smc}(s)=e^{-ms}\frac{sC^2e^{\frac{sC}{2N}}\text{Ei}\left(-\frac{sC}{2N}\right)+2N(1+4mN)}{s^2C^2\left[se^{\frac{s}{2N}}\text{Ei}\left(-\frac{s}{2N}\right)+2N\right]},
\end{equation}
where $C=1+2mN$. The prediction of Eq. \eqref{eq_avg_fracshare_s_renewal} turns out to be virtually identical (Figure \ref{fig_ibd_avg_sharing}) to the infinite-chromosome SMC expression (Eq. \eqref{eq_avg_sharing}), which can also be obtained by taking the limit $s\to 0$ (corresponding to $L\to \infty$) of Eq. \eqref{eq_avg_fracshare_s_renewal}.

\subsubsection{The variance}

\label{sect_fracshared_var}

\begin{figure}
\includegraphics[width=8cm,height=5.5cm]{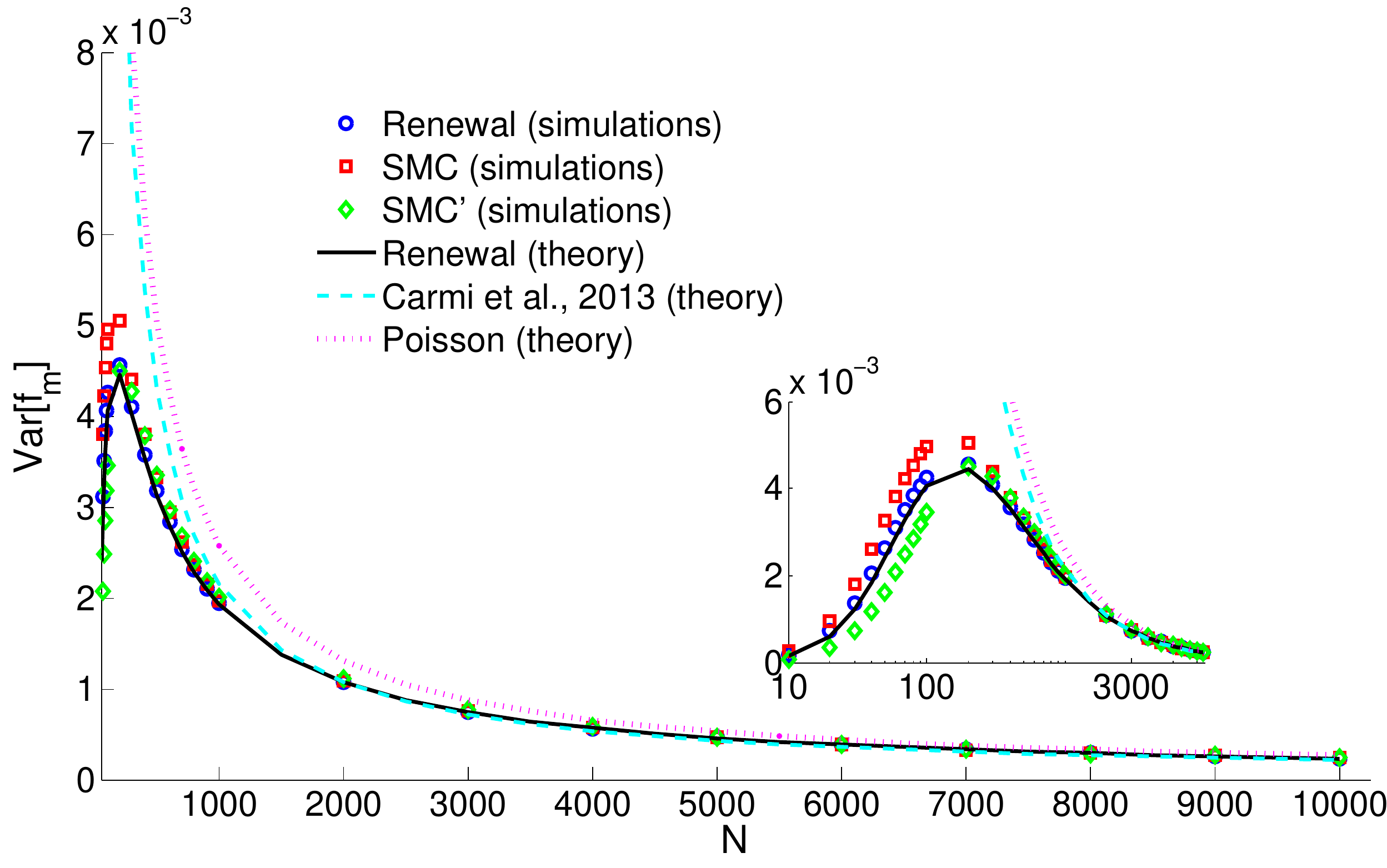}
\caption{The variance of the fraction of the chromosome found in shared segments longer than $m$, $\Var[f_m]$. Simulation details are as in Figure \ref{fig_ibd_frac_shared}. The inset zooms in on the small $N$ region. The renewal theory curve is the large $L$ expansion given in Eq. \eqref{eq_var_fracshare_lead}. The line representing \cite{HyperSharingGenetics} is from Eq. \eqref{eq_var_fracshare_carmi}, and the Poisson expression is from Eq. \eqref{eq_var_fracshare_poisson}.} \label{fig_ibd_var_sharing}
\end{figure}

The second moment of $L_m$ is given by
\begin{equation}
\tilde{\av{L_m^2}}(s)=\left.\frac{\di^2 \tP_{L_m}(u,s)}{\di u^2}\right|_{u=0}.
\end{equation}
Assuming the SMC form of $\psi(\ell)$ (Eq. \eqref{eq_seg_len_smc}), the derivatives can be taken. While the resulting expression (in $s$ space) can be numerically inverted, more insight is gained by looking at the large $L$ limit. Considering only the first order expansion in $s$ and inverting, we obtain $\lim_{L\to \infty}\av{f_m^2}=\lim_{L\to \infty}\av{f_m}^2$ or $\lim_{L\to \infty}\Var[f_m]=0$, as expected. Expanding to the next order in $s$ and inverting, we find
\begin{align}
\label{eq_var_fracshare_lead}
\Var[f_m]_{\smc}=&~\frac{\ln(1+2mN)\left[8mN\left(1+2mN-2m^3N^3\right)+1\right]}{NL(1+2mN)^4}\nonumber \\ &+
\frac{2mN\left\{8m^3N^3\ln N+mN[4mN[mN(\ln 4-1)-2]-7]-1\right\}}{NL(1+2mN)^4} \nonumber\\ &+
\frac{16m^4N^4}{(1+2mN)^4}\frac{\ln L}{NL} + O\left(\frac{\ln^2L}{L^2}\right).
\end{align}
Eq. \eqref{eq_var_fracshare_lead} is compared to simulations in Figure \ref{fig_ibd_var_sharing}, showing excellent agreement with the renewal process. For large $N$, $\Var[f_m]_{\smc}\approx [\ln\left(L/m\right)-1/2]/(NL)$. 

\cite{HyperSharingGenetics} computed the variance by approximating, for large $N$, the probability that two sites lie on shared segments, obtaining
\begin{equation}
\label{eq_var_fracshare_carmi}
\Var[f_m]\approx \frac{\ln\left(\frac{L}{m}\right)-1}{NL}.
\end{equation}
For $\ln (L/m)\gg 1$, Eq. \eqref{eq_var_fracshare_carmi} has the same limit as Eq. \eqref{eq_var_fracshare_lead}. Eq. \eqref{eq_var_fracshare_carmi} agrees well with simulations for large values of $N$ (Figure \ref{fig_ibd_var_sharing}); however, the approximation breaks down for small values of $N$.

\subsubsection{The Poisson approximation}

\label{sect_fracshare_poisson}

\cite{Palamara2012} approximated the number of shared segments longer than $m$, $n_m$, as a Poisson with mean $\av{n_m}_{\smc}=2NL(1+2mN)^2$ (Eq. \eqref{eq_ibd_avg_segnum}; see also section \ref{sect_numseg_poisson}). According to that approximation, $L_m$ can be written as a sum of $n_m$ independent random variables, each of which is distributed as $\psi_{m}(\ell)=\psi_{\smc}(\ell)/\int_m^{\infty}\psi_{\smc}(\ell)d\ell$.
To compute the distribution of $L_m$ under the Poisson approximation, $P_{\textrm{Poisson}}(L_m,L)$, it is again convenient to work in Laplace space (see also \cite{HyperSharingGenetics}). Define $\tpsi_m(u)=\int_0^{\infty}e^{-u\ell}\psi_m(\ell)d\ell$, the Laplace transform ($\ell\to u$) of $\psi_{m}(\ell)$, and denote by $\tilde{P}_{L_m,\textrm{Poisson}}(u,L)$ the Laplace transform, $L_m\to u$, of $P_{\textrm{Poisson}}(L_m,L)$.
Using the convolution theorem, given $n_m$,
\begin{equation}
\tilde{P}_{L_m,\textrm{Poisson}}(u,L|n_m)=\left[\tilde{\psi}_{m}(u)\right]^{n_m}.
\end{equation}
Since $n_m$ is assumed to be Poisson,
\begin{align}
\tilde{P}_{L_m,\textrm{Poisson}}(u,L)&=\sum_{n=0}^{\infty}e^{-\av{n_m}_{\smc}}\frac{\av{n_m}_{\smc}^n\left[\tilde{\psi}_{m}(u)\right]^{n}}{n!} \nonumber \\
&=\exp\left[-\av{n_m}_{\smc}\left(1-\tilde{\psi}_{m}(u)\right)\right].
\end{align}
This gives
\begin{align}
\label{eq_tot_sharing_laplace_poisson}
-\ln\left[\tilde{P}_{L_m,\textrm{Poisson}}(u,L)\right]/L=&~\frac{u^2 e^{\frac{u}{2N}} \text{Ei}\left[-\frac{u(1+2mN)}{2N}\right]}{2N} \\ &+\frac{e^{-mu}\left[2N\left(e^{mu}+mu-1\right)+u\right]}{(1+2mN)^2},\nonumber
\end{align}
where $\text{Ei}$ is the exponential integral function. Using Eq. \eqref{eq_tot_sharing_laplace_poisson}, $\tilde{P}_{L_m,\textrm{Poisson}}(u,L)$ can be numerically inverted \citep{deHoog}, showing (Figure \ref{fig_ibd_frac_shared}) reasonable agreement with simulation results, albeit with deviations for small values of $N$.


To compute the variance under the Poisson approximation, we redefine $\psi_m(\ell)$ as
\begin{equation}
\label{eq_def_psim_L}
\psi_m(\ell)=\frac{\psi_{\smc}(\ell)}{\int_m^{L}\psi_{\smc}(\ell)d\ell}\;\;;\;m<\ell<L,
\end{equation}
imposing an upper limit at $L$, since otherwise $\av{\ell_m^2}\to \infty$. Using the law of total variance,
\begin{align}
\Var[L_m]_{\textrm{Poisson}}&=\av{\Var[L_m|n_m]}+\Var[\av{L_m|n_m}]\\ &=
\av{n_m}_{\smc}\Var[\ell_m]+\Var[n_m]_{\smc}\av{\ell_m}^2=\av{n_m}_{\smc}\av{\ell_m^2},\nonumber
\end{align}
where we used the fact that a Poisson variable has equal mean and variance. Using Eqs. \eqref{eq_infchr_nm_general}, \eqref{eq_n0_avg_smc}, and \eqref{eq_def_psim_L},
\begin{align}
\label{eq_var_fracshare_poisson}
\Var[f_m]_{\textrm{Poisson}}&=\frac{\av{n_m}_{\smc}}{L^2}\frac{\int_m^{L}\ell^2\psi_{\smc}(\ell)d\ell}{\int_m^{L}\psi_{\smc}(\ell)d\ell}\approx\frac{2N}{L}\int_m^{L}\ell^2\psi(\ell)d\ell \nonumber\\ &=
\frac{\frac{2N(m-L)[mN(8NL+3)+3NL+1]}{(1+2mN)^2(1+2NL)^2}+\ln\left(\frac{1+2NL}{1+2mN}\right)}{NL}.
\end{align}
Here too, for large $N$, $\Var[f_m]_{\textrm{Poisson}}\approx \ln\left(L/m\right)/(NL)$, which is the same (for $\ln (L/m)\gg 1$) as the renewal theory limit (Eq. \eqref{eq_var_fracshare_lead}). Eq. \eqref{eq_var_fracshare_poisson} agrees well with simulations for large values of $N$, but breaks down already for $N\lesssim 5000$.

\section{Variable population size}

\label{sect_ibd_varsize}


Many natural populations (including humans) did not maintain a constant population size throughout their history. As we show in this section, our results are generalizable to any arbitrary variable population size, $N(t)=N_0\nu(t)$. The key insight is that all results depend on a single quantity, the PDF of segment lengths, $\psi(\ell)$. This can be seen from Eqs. \eqref{eq_infchr_n0_general}-\eqref{eq_infchr_fm_general} (the infinite-chromosome results; section \ref{sect_ibd_infinite_chr}), Eq. \eqref{eq_Pn_s_psi_final} (the distribution of the number of shared segments longer than $m$; section \ref{sect_ibd_numseg_pdf}), and Eq. \eqref{eq_Psu_fracshare_laplace} (the distribution of the fraction of the chromosome in segments longer than $m$; section \ref{sect_ibd_fracshared_pdf}). Therefore, we need only show how to compute $\psi(\ell)$ for an arbitrary $\nu(t)$. In sections \ref{sect_varsize_smc} and \ref{sect_varsize_smcprime}, we compute $\psi(\ell)$ for SMC and SMC', respectively, as well as derive the infinite-chromosome means.

\subsection{The SMC model}

\label{sect_varsize_smc}

Define $h(t)\equiv1/\nu(t)$. \cite{LiDurbin2011} derived the stationary distribution of tree heights at a recombination site (their supplementary Eq. (7)),
\begin{equation}
\label{eq_pist_smc_varsize}
\pi_{\infty}^{\smc}(t)=\frac{th(t)e^{-\int_0^th(\tau)d\tau}}{\int_0^{\infty}e^{-\int_0^{t'}h(\tau)d\tau}dt'}.
\end{equation}
Eq. \eqref{eq_pist_smc_varsize} reduces to $\pi_{\infty}^{\smc}(t)=te^{-t}$ (Eq. \eqref{eq_stat_pdf}) for a constant population size, where $h(t)=1$. For a given tree height $t$, the sequence length between recombination events is distributed exponentially with rate $2N_0t$. Therefore (see also Eq. \eqref{eq_seg_len_smc}),
\begin{align}
\label{eq_psi_ell_varsize}
\psi_{\smc}(\ell)&=\int_0^{\infty}\pi_{\infty}^{\smc}(t)\cdot 2N_0te^{-2N_0t\ell}dt  \\ &=
2N_0\frac{\int_0^{\infty}t^2h(t)e^{-\int_0^th(\tau)d\tau-2N_0t\ell}dt}{\int_0^{\infty}e^{-\int_0^{t}h(\tau)d\tau}dt} \nonumber.
\end{align}
We can now evaluate Eqs. \eqref{eq_infchr_n0_general}-\eqref{eq_infchr_fm_general} for the infinite-chromosome means. The mean segment length is
\begin{align}
\label{eq_nT_smc_varsize_derive}
\av{\ell}_{\smc}&=
2N_0\frac{\int_0^{\infty}t^2h(t)e^{-\int_0^th(\tau)d\tau}\left[ \int_0^{\infty}\ell e^{-2N_0t\ell}d\ell\right]dt}{\int_0^{\infty}e^{-\int_0^{t}h(\tau)d\tau}dt} \nonumber \\ &=
\frac{\int_0^{\infty}h(t)e^{-\int_0^th(\tau)d\tau}dt}{2N_0\int_0^{\infty}e^{-\int_0^{t}h(\tau)d\tau}dt}=\frac{1}{2N_0\int_0^{\infty}e^{-\int_0^{t}h(\tau)d\tau}dt}.
\end{align}
Hence (see Eq. \eqref{eq_infchr_n0_general}),
\begin{equation}
\label{eq_nT_smc_varsize}
\av{n_0}_{\smc}=2N_0L\int_0^{\infty}e^{-\int_0^{t}h(\tau)d\tau}dt.
\end{equation}
Note that we implicitly assumed that that $\lim_{t\to \infty}\nu(t)<\infty$. Eq. \eqref{eq_nT_smc_varsize} can also be obtained using Corollary 3 in \cite{LiDurbin2011}. For the mean number of segments longer than $m$, we obtain, using techniques similar to those used in Eq. \eqref{eq_nT_smc_varsize_derive},
\begin{equation}
\label{eq_nm_smc_varsize}
\av{n_m}_{\smc}=2N_0L\int_0^{\infty}th(t)e^{-\int_0^t h(\tau)d\tau-2N_0mt}dt.
\end{equation}
Finally,
\begin{equation}
\label{eq_fm_smc_varsize}
\av{f_m}_{\smc}=\int_0^{\infty}h(t)e^{-\int_0^t h(\tau)d\tau-2N_0mt}(1+2N_0mt)dt.
\end{equation}
Eq. \eqref{eq_fm_smc_varsize} was also derived by \cite{Palamara2012}. It can be verified that substituting $h(t)=1$ in Eqs. \eqref{eq_nT_smc_varsize}, \eqref{eq_nm_smc_varsize}, and \eqref{eq_fm_smc_varsize}, we recover the results of section \ref{sect_ibd_smc_inf_chr} (Eqs. \eqref{eq_n0_avg_smc}, \eqref{eq_ibd_avg_segnum}, and \eqref{eq_avg_sharing}, respectively).

\subsection{The SMC' model}

\label{sect_varsize_smcprime}

For SMC', we need to recompute $q_{\smcp}(t|s)$, the probability that the new tree height at a recombination site is $t$, given that the previous height was $s$ (see Eq. \eqref{eq_transition_pdf_smcprime_raw}),
\begin{equation}
\label{eq_transition_pdf_smcprime_raw_varsize}
q_{\smcp}(t|s) = \begin{cases}
\int_0^s\frac{1}{s}\left[\int_{t_r}^{s}h(t_c)e^{-2\int_{t_r}^{t_c}h(\tau)d\tau}dt_c\right]dt_r & t=s, \\
\int_0^t\frac{1}{s}h(t)e^{-2\int_{t_r}^{t}h(\tau)d\tau}dt_r & t<s, \\
\left[\int_0^s\frac{1}{s}e^{-2\int_{t_r}^{s}h(\tau)d\tau}dt_r\right] h(t)e^{-\int_{s}^{t}h(\tau)d\tau} & t>s.
\end{cases}
\end{equation}
Eq. \eqref{eq_transition_pdf_smcprime_raw_varsize} is explained similarly to Eq. \eqref{eq_transition_pdf_smcprime_raw}, once we recognize that coalescence occurs at (absolute) time $t$ at rate $h(t)$, and that the probability of \emph{no} coalescence between $[s,t]$ is $e^{-\int_s^th(\tau)d\tau}$ \citep{GriffithsTavare94}. It can be shown that Eq. \eqref{eq_transition_pdf_smcprime_raw_varsize} is normalized ($\int_0^tq_{\smcp}(t|s)dt=1$), and that, as in the case of a constant population size (section \ref{sect_ibd_process_smcprime}), the stationary distribution of tree heights, $\pi_{\infty}^{\smcp}(t)$, is identical to that of SMC and is given by Eq. \eqref{eq_pist_smc_varsize}. It can also be shown that at stationarity, the new tree has equal probabilities to be either taller, shorter, or equal to the previous tree, as we have seen for a constant population size (section \ref{sect_ibd_smcprime_avg_sharing}).

As in section \ref{sect_ibd_smcprime_avg_sharing}, we define a chain with probabilities $q_{\smcpseg}(t|s)=q_{\smcp}(t|s)/[1-q_{\smcp}(s|s)]$ (as in Eq. \eqref{eq_qpp_def}), whose stationary distribution, $\pi_{\infty}^{\smcpseg}(t)$, is the distribution of tree heights at segment ends. Using the marginal distribution of tree heights at random sites \citep{GriffithsTavare94},
\begin{equation}
\label{eq_pc_var}
P_c(t)=h(t)e^{-\int_0^t h(\tau)d\tau},
\end{equation}
and a detailed balance argument, it can be shown that
\begin{equation}
\label{eq_pistat_smcp_segend_varsize}
\pi_{\infty}^{\smcpseg}(t)= \frac{P_c(t)\lambda(t)}{\int_0^{\infty}P_c(t)\lambda(t)dt},
\end{equation}
where
\begin{align}
\label{eq_smcprime_lambda_varsize}
\lambda(t)&=2N_0t[1-q_{\smcp}(t|t)] \\ &=
2N_0t\left[1-\frac{1}{t}\int_0^t\!\!\int_{t_r}^{t}h(t_c)e^{-2\int_{t_r}^{t_c}h(\tau)d\tau}dt_cdt_r\right] \nonumber \\ &= N_0\left[t+e^{-2\int_0^t h(\tau)d\tau}\int_0^t e^{2\int_0^{t'}h(\tau)d\tau}dt'\right].
\end{align}
The distribution of segment lengths is then given by (see also Eq. \eqref{eq_psi_smcprime_integral}; section \ref{sect_ibd_process_smcprime})
\begin{align}
\label{eq_smcprime_varsize_psiell_general}
\psi_{\smcp}(\ell)&=\int_0^{\infty}\pi_{\infty}^{\smcpseg}(t)\lambda(t)e^{-\lambda(t)\ell}dt \nonumber \\ &=
\frac{\int_0^{\infty} P_c(t)\lambda^2(t)e^{-\lambda(t)\ell}dt}{\int_0^{\infty}P_c(t)\lambda(t)dt}.
\end{align}
Note that Eq. \eqref{eq_smcprime_varsize_psiell_general} depends solely on $N(t)$, and as expected, the distribution of $\rho=2N_0\ell$ is independent of $N_0$.

We now derive the infinite-chromosome means (section \ref{sect_ibd_infinite_chr}). The mean segment length is 
\begin{equation}
\av{\ell}_{\smcp}=\int_0^{\infty}\ell\psi_{\smcp}(\ell)d\ell=\left[\int_0^{\infty}P_c(t)\lambda(t)dt\right]^{-1},
\end{equation}
where we used the fact that $\int_0^{\infty}P_c(t)dt=1$. Using Eq. \eqref{eq_infchr_n0_general} and after some algebra, 
\begin{align}
\label{eq_n0_smcprime_varsize}
\av{n_{0}}_{\smcp}&=L\int_0^{\infty}P_c(t)\lambda(t)dt \nonumber \\
&=N_0L\int_0^{\infty}h(t)e^{-\int_0^t h(\tau)d\tau}\left[t+e^{-2\int_0^t h(\tau)d\tau}\int_0^t e^{2\int_0^{t'}h(\tau)d\tau}dt'\right]dt \nonumber \\ 
&=\frac{4N_0L}{3}\int_0^{\infty}e^{-\int_0^t h(\tau)d\tau}dt.
\end{align}
This is, as expected, exactly $2/3$ of the number of recombination events (Eq. \eqref{eq_nT_smc_varsize}).

Using Eqs. \eqref{eq_pc_var}, \eqref{eq_smcprime_varsize_psiell_general}, and \eqref{eq_n0_smcprime_varsize}, we can write
\begin{align}
\label{eq_smcprime_varsize_psiell_simple}
\psi_{\smcp}(\ell)&= \frac{\int_0^{\infty} P_c(t)\lambda^2(t)e^{-\lambda(t)\ell}dt}{\av{n_{0}}_{\smcp}/L}\nonumber \\ &=\frac{\int_0^{\infty} h(t)\lambda^2(t)e^{-\int_0^th(\tau)d\tau-\lambda(t)\ell}dt}{\frac{4N_0}{3}\int_0^{\infty}e^{-\int_0^t h(\tau)d\tau}dt}.
\end{align}
The mean number of segments longer than $m$ is
\begin{align}
\av{n_{m}}_{\smcp}&=\av{n_{0}}_{\smcp}\int_m^{\infty}\psi_{\smcp}(\ell)d\ell\nonumber \\ &
=L\int_0^{\infty}P_c(t)\lambda(t)e^{-\lambda(t)m}dt.
\end{align}
Finally, the mean fraction of the chromosome in segments longer than $m$ is
\begin{align}
\av{f_{m}}_{\smcp}&=\frac{\av{n_{0}}_{\smcp}}{L}\int_m^{\infty}\ell\psi_{\smcp}(\ell)d\ell\nonumber \\ &=\int_0^{\infty}P_c(t)e^{-\lambda(t)m}[1+\lambda(t) m]dt.
\end{align}
It can be verified that all the results of this section reduce to the SMC results (section \ref{sect_varsize_smc}) for $\lambda(t)=2N_0t$ and to the constant population size results (section \ref{sect_ibd_smcprime_avg_sharing}) for $h(t)=1$.

\section{Summary and discussion}

\label{sect_discussion}

In summary, we introduced a general framework for the IBD process in Markovian approximations of the coalescent with recombination (SMC and SMC'), as well as a new renewal approximation, in which tree heights on both sides of a recombination site are independent (section \ref{sect_ibd_process}). We showed how previous results for the mean number of segments and the mean shared sequence length in SMC emerge naturally from our framework in the infinite-chromosome limit; we then derived these quantities under SMC' (section \ref{sect_ibd_infinite_chr}).  Using renewal theory, we derived expressions for the distributions of the number of shared segments (section \ref{sect_ibd_numseg_pdf}) and the fraction of the chromosome in shared segments (section \ref{sect_ibd_fracshared_pdf}). Finally, we generalized our results to populations with variable size (section \ref{sect_ibd_varsize}). 

Our main contributions are \begin{inparaenum}[\itshape a\upshape)]
\item providing a unified framework for the IBD process, depending exclusively on a single distribution (that of segment lengths), in which previous and new results are coherently derived and easily generalized; 
\item new results for SMC': the distribution of tree heights at recombination sites (both conditional on the previous tree and at stationarity), the stationary distribution of tree heights at segment ends, the distribution of segment lengths, the mean number of shared segments, and the mean fraction of the chromosome in shared segments; and
\item introducing a novel renewal approximation, under which distributions of key quantities were obtained.
\end{inparaenum}

Our results rely on a number of simplifying assumptions, beyond the standard postulates of coalescent theory. First, our model considers segments shared between haploid chromosomes and does not incorporate any model for shared segments detection errors. In reality, genotyping errors, recent mutations, and phase uncertainty do not allow the confident detection of short segments, although this is partly remedied by our theory being entirely specified in terms of a length cutoff ($m$), which can be tuned for the quality of the data under examination. Next, when computing distributions, we assumed that sharing between each pair of chromosomes is independent, whereas in practice, scans for IBD search for shared segments between all pairs in a cohort. Indeed, as studied in detail by \cite{HyperSharingGenetics}, while sharing between two pairs in a cohort is only weakly dependent, the cumulative effect increases the observed variance of the amount of overall sharing. Therefore, more work will be needed to understand the distribution of IBD sharing within a cohort. Finally, we derived all results for a single chromosome. To apply the results genome-wide, we must assume inter-chromosome independence, which may not be well justified for the very recent past \citep{Wakeley_pedigree}.

Turning to the quality of the renewal approximation itself, we verified using simulations that for chromosome-wide properties (e.g. the total number of segments), the renewal results are indistinguishable from SMC. We do, however, expect small deviations for very short chromosomes and for very small populations (e.g., see Figure \ref{fig_ibd_var_sharing}), when segments are few and long compared to the chromosome length and the distribution of tree heights does not reach stationarity. We also note that as opposed to SMC and SMC' \citep{SMCprime}, the renewal approximation introduces an asymmetry between the two ends of the chromosome: while the segment at the left end has distribution $\psi(\ell)$, the segment at the right end has the distribution of the `age' of the process (see \cite{KarlinTaylor} for more details). As we explained in section \ref{sect_ibd_process_intro}, the number of segments is typically so large that this has a negligible effect. However, one can also formulate a \emph{stationary} renewal process, which begins at coordinate $-\infty$, while observations begin at the origin \citep{KarlinTaylor}. With some effort, we could rederive all results under the stationary process (not shown).

Our results have consequences for demographic inference. Current approaches rely on the assumption that recombination events terminate shared segments, as in SMC \citep{Palamara2012,RalphCoop}. Using our results, the more accurate SMC' can now be used, particularly for small populations. The distribution of the number of shared segments is also expected to be useful, as we briefly demonstrated (Figure \ref{fig_ibd_inference}). The case we studied is simple, and would have been easily solved by other methods (e.g., \cite{Palamara2012,HyperSharingGenetics}). Nevertheless, our approach has the attractive feature of providing a maximum-likelihood estimator (under the assumptions discussed above). Of course, for either large populations or for the very remote past, long IBD segments are scarce and our method, like any other IBD-based estimator, will have limited power.

Another drawback of our method is that it requires a numerical Laplace transform inversion, and for complex demographies, even the Laplace space solution will have to be numerically computed. Nevertheless, computationally, this is not very different from any method based on results specified as integrals or sums. The inverse transform (at least for the distribution of the number of shared segments) was simple to compute and reliable, as we validated by simulations (e.g., Figure \ref{fig_ibd_numseg_dist}), as well as by comparing a number of inversion methods (not shown). Running time was reasonably short, at $\approx 2.5$ seconds for each $N$ on a standard machine. We anticipate that using the results for the fraction of the chromosome in shared segments (section \ref{sect_ibd_fracshared_pdf}) will have more limited applications, due to the need for a double Laplace transform inversion. But we also note that, as we showed in sections \ref{sect_numseg_mean}, \ref{sect_numseg_var}, \ref{sect_fracshared_mean}, and \ref{sect_fracshared_var}, standard Laplace transform techniques allow insight into the moments of the examined distributions. The Laplace transform method is ideal for problems of Markovian evolution in time or sequence space that are otherwise difficult (e.g., \cite{LohseLaplace}), and is therefore expected to be of future interest in population genetics.

We foresee a number of future directions and potential extensions. First, it would be useful (e.g., for demographic inference) to have analytical forms for simple non-constant demographies, such as exponential expansions and bottlenecks. Second, while we provided an equation for $\psi_{\smcp}(\ell)$, the PDF of segment lengths in SMC' (Eq. \eqref{eq_psi_smcprime_integral}), we did not investigate the corresponding renewal approximation (beyond the infinite-chromosome means), which should be feasible, since all of our renewal-based results are given in terms of a general segment length distribution. This is expected to rise in importance with the increasing popularity of SMC' (e.g., \cite{HarrisNielsen}) and the emerging understanding that it provides a much better approximation to the coalescent with recombination than SMC (e.g., \cite{Hobolth2014}). Another potential future application is pedigree reconstruction using IBD segments \citep{Huff_ERSA,Henn2012}. For example, for (half-) cousins separated by $2k$ meioses, the segment length distribution will be a superposition of an exponential with rate $2k$, with probability $2^{-2k}$, and $\psi_{\smc}(\ell)$ or $\psi_{\smcp}(\ell)$ otherwise (Eqs. \eqref{eq_seg_len_smc} and \eqref{eq_psi_smcprime_integral}, respectively). Finally, a challenging extension will be to sharing between more than two chromosomes. 
Potentially interesting applications are awaiting, as methods for the detection of such segments have been developed \citep{DASH,Moltke_MultipleIBD,He_MultipleIBD}, and the resulting information is expected to improve the accuracy of demographic inference, natural selection detection, and disease mapping.  

\section*{Acknowledgements}

We thank Asger Hobolth for commenting on the manuscript and for pointing out a number of arguments regarding Markov chains reversibility and detailed balance. S. C. thanks Eli Barkai, whose lecture notes on renewal theory have been heavily consulted, and the Human Frontier Science Program for financial support. I. P. thanks NIH grant 1R01MH095458-01A1.

\section*{References}

\bibliographystyle{elsarticle-harv}
\bibliography{../Renewal}

\appendix


\section{Full expressions for SMC' results}
\label{sect_smcprime_appendix}

In this section, we provide full expressions for a number of SMC' quantities that were expressed as integrals in the main text.

The full expression for the distribution of segment lengths (Eq. \eqref{eq_psi_smcprime_integral}, section \ref{sect_ibd_process_smcprime}) is
\begin{align}
\label{eq_smcprime_psi_ell_full}
\psi_{\smcp}(\ell)&=\frac{\int_0^{\infty}e^{-t}\lambda^2(t)e^{-\lambda(t)\ell}dt}{\int_0^{\infty}e^{-t}\lambda(t)dt} \\
&=3e^{-\frac{q}{2}(1+2\pi i)}q^{-\frac{q+1}{2}}\left[64\ell q \Gamma\left(\frac{1-q}{2}\right)^2\right]^{-1} \times
\nonumber \\ &\quad \left\{\pi^2q^2e^{i\pi q}q^{\frac{q+1}{2}}\sec^2\left(\frac{\pi q}{2}\right)\times \right.\nonumber \\ &\quad\left.\left[4\, _2\tilde{F}_2\left(\frac{q+1}{2},\frac{q+1}{2};\frac{q+3}{2},\frac{q+3}{2};\frac{q}{2}\right)\right.\right.\nonumber \\ &\quad \left.\left.-~(q+1)^2\,_2\tilde{F}_2\left(\frac{q+3}{2},\frac{q+3}{2};\frac{q+5}{2},\frac{q+5}{2};\frac{q}{2}\right)\right.\right. \nonumber \\ &\quad \left.\left. +~4\Gamma \left(\frac{q+1}{2}\right) \, _3\tilde{F}_3\left(\frac{q+1}{2},\frac{q+1}{2},\frac{q+1}{2};\frac{q+3}{2},\frac{q+3}{2},\frac{q+3}{2};\frac{q}{2}\right)\right]\right.\nonumber \\&\quad\left.+~i2^{\frac{q+3}{2}} e^{\frac{i\pi q}{2}}\Gamma\left(\frac{1-q}{2}\right)\times \right. \nonumber \\ &\quad\left. \left[\Gamma\left(\frac{1-q}{2}\right)\left[q^2\Gamma \left(\frac{q+1}{2},-\frac{q}{2}\right)+4q\Gamma\left(\frac{q+3}{2},-\frac{q}{2}\right)+4\Gamma\left(\frac{q+5}{2},-\frac{q}{2}\right)\right]\right.\right. \nonumber \\ &\quad \left. \left.-~\pi\sec\left(\frac{\pi q}{2}\right)\left(4q^2+6q+3\right)\right]\right\}\nonumber,
\end{align}
where $q=N\ell$, $\Gamma$ (with two arguments) is the incomplete Gamma function, and $_a\tilde{F}_b$ is the regularized generalized hypergeometric function \citep{MathWorld}. See simulation results in Figure \ref{fig_len_pdf}. Note that $\psi_{\smcp}(\ell)$ is necessary real (and similarly below).

The full expression for the mean number of shared segments longer than $m$ (Eq. \eqref{eq_nm_smcprime_integral}, section \ref{sect_ibd_smcprime_avg_sharing}) is
\begin{align}
\label{eq_smcprime_avg_nm_full}
\av{n_{m}}_{\smcp}&=L\int_0^{\infty}\lambda(t)e^{-t-\lambda(t)m}dt \nonumber \\ &=LN i\frac{(\frac{-eM}{2})^{-\frac{M}{2}}}{2\sqrt{2M}} \times \\ & \quad\left\{
\Gamma \left(\frac{M+1}{2},-\frac{M}{2}\right)+\frac{2}{M}\Gamma\left(\frac{M+3}{2},-\frac{M}{2}\right)\right.\nonumber \\ & \quad\left. +~\Gamma\left(\frac{M+1}{2}\right) \left[\psi ^0\left(\frac{M+1}{2}\right)-2-i\pi-\ln \frac{M}{2}-\frac{1}{M}\right]\right.\nonumber \\ & \quad\left.
-~G_{2,3}^{3,0}\left(-\frac{M}{2}\left|
\begin{array}{c}
 1,1 \\
 0,0,\frac{M+1}{2} \\
\end{array}
\right.\right)\right\}, \nonumber
\end{align}
where $M=mN$, $G$ is the Meijer G-function \citep{MathWorld} and $\psi^0$ is the digamma function.

The full expression for the mean fraction of the chromosome in segments longer than $m$ (Eq. \eqref{eq_avg_sharing_smcprime_integral}, section \ref{sect_ibd_smcprime_avg_sharing}) is
\begin{align}
\label{eq_avg_sharing_smcprime}
\av{f_{m}}_{\smcp}&=\int_0^{\infty}e^{-t-\lambda(t)m}[1+\lambda(t) m]dt\nonumber \\ &=
\frac{(\frac{-eM}{2})^{-\frac{M}{2}}}{2\sqrt{2M}} \left\{\frac{M^{3/2}}{\sqrt{2}} G_{2,3}^{3,0}\left(-\frac{M}{2}\left|
\begin{array}{c}
 \frac{1}{2},\frac{1}{2} \\
 -\frac{1}{2},-\frac{1}{2},\frac{M}{2} \\
\end{array}
\right.\right)\right.  \\ &\left.+~i(M+2)\Gamma \left(\frac{M+1}{2},-\frac{M}{2}\right)+2i\Gamma \left(\frac{M+3}{2},-\frac{M}{2}\right)\right. \nonumber \\ &\left.+~iM\Gamma \left(\frac{M+1}{2}\right) \left[\psi^0\left(\frac{M+1}{2}\right)-\ln\frac{M}{2}-2-i\pi-\frac{3}{M}\right]\right\}\nonumber,
\end{align}
where $M=mN$. See simulations results in Figure \ref{fig_ibd_avg_sharing}.

\section{Full expression for the renewal theory results}

\label{sect_renewal_appendix}

In the renewal approximation to SMC, the distribution of the number of segments longer than $m$, in Laplace space (Eq. \eqref{eq_Pn_s_psi_final}; section \ref{sect_ibd_numseg_dist_derive}), is
\begin{align}
\label{eq_Pn_s_smc}
\tP(n_m=k,s)=&~ C^{-2}2^{3-n}N^2e^{-ms}\left[se^{\frac{s}{2N}}\text{Ei}\left(-\frac{s}{2N}\right)+2N\right]\\ & \times \left\{s^2e^{\frac{s}{2N}} \left[E_1\left(\frac{s}{2N}\right)-E_1\left(\frac{sC}{2N}\right)\right]+2D-2Ns\right\}^{-2} \nonumber \\ & \times \left\{\frac{s^2e^{\frac{s}{2N}}\text{Ei}\left(-\frac{sC}{2N}\right)+2D}{\frac{s^2}{2}e^{\frac{s}{2N}}\left[\Gamma\left(0,\frac{s}{2N}\right)-\Gamma\left(0,\frac{sC}{2N}\right)\right]-Ns+D }\right\}^{n-1}\nonumber
\end{align}
for $k>0$ and
\begin{equation}
\tP(n_m=0,s)=\left\{\frac{4N^2C^{-1}}{e^{ms}C\left[2N-se^{\frac{s}{2N}}\left[E_1\left(\frac{s}{2N}\right)-E_1\left(\frac{sC}{2N}\right)\right]\right]-2N}+s\right\}^{-1}
\end{equation}
for $k=0$, where $C=1+2mN$, $D=Ne^{-ms}(sC-2N)/C^2$, and $E_1$ is related to the exponential integral function ($E_1(x)=-E_i(-x)$) \citep{MathWorld}.

The distribution of the fraction of the chromosome in segments longer than $m$, in Laplace space (Eq. \eqref{eq_Psu_fracshare_laplace}, section \ref{sect_ibd_frac_shared_derive}), is
\begin{equation}
\label{eq_Psu_fracshare_smc}
\tP_{L_m}(u,s)=A/(1-B),
\end{equation}
where $A$ is given by
\begin{align}
4C^2A=&~\frac{4e^{-mr}}{r}-\frac{4e^{-ms}}{s}+\frac{4}{s}+\frac{2\left(1-e^{-ms}\right)}{N}\\&+\frac{e^{-mr}}{N^2r}\left[C^2r^2e^{\frac{Cr}{2N}}\text{Ei}\left(-\frac{Cr}{2N}\right)+2N(Cr-2N)\right] \nonumber\\ &+\frac{4e^{-ms}}{Ns}\left\{\frac{s}{2}\left(e^{ms}-1\right)+2m^2N^2se^{m s}+N\left[e^{ms}(2ms-1)+1-ms\right]\right\}\nonumber\\&+\frac{sC^2e^{\frac{s}{2N}}}{N^2}\left[\Gamma\left(0,\frac{sC}{2N}\right)-\Gamma\left(0,\frac{s}{2N}\right)\right],\nonumber
\end{align}
$B$ is given by
\begin{align}
4N^2C^2B=&~4N^2\left[4m^2N^2+2mN(2-ms)+1-2ms+e^{-ms}\left(ms-1\right)\right] \nonumber \\
&+2Ns\left(e^{-ms}-1\right) +C^2s^2e^{\frac{s}{2N}}\left[\Gamma\left(0,\frac{s}{2N}\right)-\Gamma\left(0,\frac{sC}{2N}\right)\right] \nonumber \\ &-e^{-mr}\left[C^2r^2e^{\frac{Cr}{2N}}\text{Ei}\left(-\frac{Cr}{2N}\right)+2N(Cr-2N)\right],
\end{align}
$C=1+2mN$, and $r=s+u$.



\end{document}